\documentclass{aa}
\usepackage{graphics}
\usepackage{graphicx}
\thesaurus{02(11.03.1; 11.19.7; 12.03.3; 12.07.1; 13.25.2)}
\begin{document}
\title{The giant luminous arc Statistics.}
\subtitle{II. spherical lens models based on {\sl ROSAT HRI}\ data.}
\author{Kohji Molikawa\inst{1} \and Makoto Hattori\inst{1} \and
Jean-Paul Kneib\inst{2} \and Kazuyuki Yamashita\inst{3}}
\institute{Astronomical Institute, Graduate School of Science,
T\^ohoku University, Sendai, Miyagi 980-8578, Japan \and
Observatoire Midi-Pyr\'en\'ees, Laboratoire d'Astrophysique,
14 Av. E. Belin, 31400 Toulouse, France \and
Information Processing Center, Chiba University, Inage, Chiba 263-8522, Japan}
\offprints{K. Molikawa}
\date{Received / Accepted }
\maketitle
\begin{abstract}
We present {\sl ROSAT HRI}\ X-ray observations of 
all the galaxy clusters in the Le F\`evre et al. arc survey sample
in order to study the spatial distribution of the intra-cluster medium (ICM),
and examine the expected number of giant luminous arcs for the sample
using two spherically symmetric lens models
constrained by our X-ray data.
{\it Isothermal $\beta$ model} assumes that the ICM is isothermal 
and is in the hydrostatic equilibrium. {\it ENF98-NFW model}
assumes that the `universal' dark matter halo profile 
proposed by Navarro, Frenk \& White is a valid description 
of the underlying dark matter distribution.
Adopting the result of N-body/gas-dynamical simulations by Eke, Navarro \& Frenk,
dark matter distribution in the {\it ENF98-NFW model} can be constrained by the 
X-ray surface brightness distribution of the ICM.
The expected number of giant luminous arcs in the sample 
is then calculated taking into account both detection conditions in 
the arc survey and the evolution of source galaxies.
We find that the {\it isothermal $\beta$ model}
 cannot reproduce the observed number 
of giant luminous arcs even allowing uncertainties in the source 
galaxy model.
The {\it ENF98-NFW model} displays good agreement
in number of giant luminous arcs.
However, some clusters have their virial temperature
3--4 times higher than their X-ray temperature
obtained from spectral data or from the $L_{\rm X}-T$ relation.
Thus, we conclude that 
both spherical models which are consistent with all the available X-ray data 
cannot reproduce the observed number of giant luminous arcs in the sample.
To solve this discrepancy we believe that
the giant luminous arc statistics
will need to take properly into account
`irregularity' in the mass distribution in each cluster.
\end{abstract}
%
%
\section{Introduction}

A number of Giant luminous arcs (GLAs) have now been detected in
many distant clusters of galaxies (e.g.\ Soucail et al.\ \cite{Soucail87}; 
Lynds \& Petrosian \cite{Lynds86}; Fort \& Mellier \cite{Fort94} for a review).
Their spectroscopic confirmations as lensed normal distant galaxies
(e.g.\ Soucail et al.\ \cite{Soucail88})
lead to exciting new cosmological developments in this decade.

Probing the mass distribution of distant clusters is one of avenues of these
developments (e.g.\  Narayan \& Bartelmann \cite{Narayan95}; 
Hattori et al.\ \cite{Hattori99}; Umetsu et al.\ \cite{Umetsu99}).
Early analyses pointed out discrepancy of a factor 2 -- 3
between cluster masses derived from strong lensing and
standard X-ray analysis of the ICM
(e.g.\ Loeb \& Mao \cite{Loeb94}; Miralda-Escud\'e \& Babul \cite{Miralda95}).
Several possible solutions are proposed:
(i) Loeb \& Mao \cite{Loeb94} proposed that non-thermal pressure 
due to equipartition magnetic field and turbulent may play an important role on supporting the ICM.
(ii) A multi-phase model of the ICM in the central part of clusters -- indicated 
by the cooling flow model (Allen et al.\ \cite{Allen96}) -- 
increases by a factor 2--3 the mass deduced from X-ray  compared
to standard (single-phase) isothermal model (Allen \cite{Allen98}).
(iii) Leaving from "inappropriate modeling of the cluster mass distribution 
due to neglecting the contributions by substructure and member galaxies,
which leads overestimation of the cluster mass from strong lensing events 
(e.g.\ Kneib et al.\ \cite{Kneib93}; Kneib et al.\ \cite{Kneib95};
Kneib et al.\  \cite{Kneib96};
Hattori et al.\ \cite{Hattori98}).
One can expect that the next generation X-ray telescopes (Chandra, XMM, Astro-E)
will provide definite answer of the first two possible solutions.
The third one seems, however, the closest one to the reality (Hattori et al. \cite{Hattori99}).

Instead of these detailed analyses for individual clusters,
Statistics of GLAs constrains the average properties of cluster mass distribution.
The statistics of GLAs is consist in counting the number of GLAs in a well defined cluster sample
and constrains the average properties of cluster mass distribution.
Using spherically symmetric mass distribution models, 
Wu \& Hammer (\cite{Wu93}) showed
that the predicted number of GLAs was critically dependent
on the degree of the central mass concentration of clusters.
Miralda-Escud\'e (\cite{Miralda93a}) examined arc statistics
with elliptical lens models and concluded that elliptical mass
distribution did not change the above conclusion drastically.
Using numerically simulated model for clusters,
Bartelmann and his collaborators (Bartelmann \& Weiss \cite{Bartelmann94}; 
Bartelmann et al.\ \cite{Bartelmann95}, henceforth BSW95)
showed that their asymmetric cluster mass distribution 
(e.g.\ substructure) increased the expected number of GLAs.
In a further study, Bartelmann et al.\ (\cite{Bartelmann98}) examined
the dependence of arc statistics on cosmological parameters.
They concluded that the open cold dark matter model
gave the largest number of giant arcs compared to other cosmology, in
particular the Einstein-de Sitter universe
and that the open cold dark matter model was likely
to be the only model which could match current observations.

The first well-defined GLA survey was conducted by Le F\`evre et al
(\cite{LF94}, henceforth LF94).
Hattori et al.\ (\cite{Hattori97}, henceforth Paper I) proposed 
a method that predicted the number of GLAs taking into account
both detection conditions in the arc survey
and the evolution of source galaxies.
They applied their method to  the LF94 arc survey sample using 
the {\sl Einstein} X-ray data to calibrate the mass distribution - 
assuming the hydrostatic equilibrium and the spherical symmetry.
The conclusion of Paper I is
that the expected number of GLAs deduced from the X-ray analysis 
was inconsistent with the observed one. 
Hamana \& Futamase (\cite{Hamana97}, henceforth HF97)
examined the GLA statistics taking into account the evolution of the luminosity function
found by the Canada France Redshift Survey (Lilly et al.\ \cite{Lilly95}).
They showed that the observed evolution in the galaxy luminosity function increased the
expected GLA number by a factor of 2 -- 3 at most.
Although the $L_{\rm X}-\sigma$ relation adopted by them and by Wu \& Hammer (\cite{Wu93})
appeared to be incorrect when compared with recent results
(e.g.\ Mushotzky \& Scharf \cite{Mushotzky97}; Markevitch \cite{Markevitch98}),
we can study how sensitive the GLA statistics is 
on the evolution of the galaxy luminosity function.

This paper is the second in a series on the giant luminous arc statistics
with the LF94 arc survey sample.
In this paper we re-visit the LF94 sample 
with higher quality {\sl ROSAT HRI}\ data
because no precise measurements of the ICM spatial distribution in the LF94 sample clusters 
were available in Paper I.
We assume that the mass distribution in clusters are spherically symmetric.
This paper is organized as follows.
In \S 4, we describe the lensing properties of mass models.
We discuss the GLA statistics and its implication in \S 5,
and summarize our conclusion in \S 6.
Notes on individual clusters,
as well as contour plot and radial profile of each cluster,
are given in the appendix.
Throughout this paper, we adopt a Hubble constant 
of $H_{0} = 50 h_{50}~{\rm km~s^{-1}~Mpc^{-1}}$, 
the present density parameter of ${\Omega_{\rm m}}_0 = 0.3$,
and the present cosmological parameter of ${\Omega_{\Lambda}}_0 = 0.7$.

%
%
\begin{table*}[t]
\caption[]{The log of {\sl ROSAT HRI}\ observations of the sample clusters.}
\label{Log}
\begin{tabular}{lcccccc}
\hline\hline\\
Cluster name & Redshift &
 Pointing R.A. & Pointing Dec. &
 Observation date & Exposure & Column density \\
 & & (J2000.0) & (J2000.0) & & $[$ ksec. $]$ & $[\:10^{21}~{\rm cm}^{-2}\:]$\\
\hline\\
\object{MS 0015.9$+$1609}  & $0.546$ & 
$00^{\rm h}18^{\rm m}33\fs6$ & 
$+16\degr 26\arcmin 24\arcsec$ & 
95 Jan 09 -- 95 Jul 05 & $76.1$ & $0.41$ \\
\object{MS 0302.7$+$1658}  & $0.426$ & 
$03^{\rm h}05^{\rm m}33\fs6$ & 
$+17\degr 10\arcmin 12\arcsec$ & 
95 Aug 19 -- 95 Aug 27 & $33.8$ & $1.07$ \\
\object{MS 0353.6$-$3642}  & $0.320$ & 
$03^{\rm h}55^{\rm m}31\fs2$ & 
$-36\degr 33\arcmin 36\arcsec$ & 
94 Jul 19 -- 94 Aug 21 & $22.2$ & $0.12$ \\
\object{MS 0451.5$+$0250}    & $0.202$ & 
$04^{\rm h}54^{\rm m}09\fs6$ & 
$+02\degr 55\arcmin 12\arcsec$ & 
94 Mar 05 -- 94 Mar 06 & $12.7$ & $0.78$ \\
\object{MS 0735.6$+$7421}  & $0.216$ & 
$07^{\rm h}41^{\rm m}50\fs4$ & 
$+74\degr 15\arcmin 00\arcsec$ & 
92 Mar 14 -- 92 Apr 21 & $27.0$ & $0.41$ \\
\object{MS 1006.0$+$1202}  & $0.221$ & 
$10^{\rm h}08^{\rm m}45\fs6$ & 
$+11\degr 48\arcmin 00\arcsec$ & 
96 Jun 03 -- 96 Dec 06 & $22.4$ & $0.37$ \\
\object{MS 1008.1$-$1224}  & $0.301$ & 
$10^{\rm h}10^{\rm m}33\fs6$ & 
$-12\degr 39\arcmin 36\arcsec$ & 
94 May 16 -- 94 Jun 17 & $68.7$ & $0.73$ \\
\object{MS 1224.7$+$2007}  & $0.327$ & 
$12^{\rm h}27^{\rm m}14\fs4$ & 
$+19\degr 51\arcmin 00\arcsec$ & 
94 Jun 20 -- 96 Jun 05 & $37.1$ & $0.29$ \\
\object{MS 1333.3$+$1725}  & $0.460$ & 
$13^{\rm h}35^{\rm m}48\fs0$ & 
$+17\degr 09\arcmin 36\arcsec$ & 
97 Jul 11 -- 97 Jul 24 & $54.4$ & $0.18$ \\
\object{MS 1358.4$+$6245}  & $0.328$ & 
$13^{\rm h}59^{\rm m}50\fs3$ & 
$+62\degr 31\arcmin 12\arcsec$ & 
91 Nov 05 -- 93 May 15 & $29.2$ & $0.19$ \\
MS 1455.0$+$2232  & $0.259$ & 
$14^{\rm h}57^{\rm m}14\fs4$ & 
$+22\degr 20\arcmin 24\arcsec$ & 
92 Jan 11 -- 94 Jul 08 & $14.8$ & $0.33$ \\
\object{MS 1512.4$+$3647}  & $0.372$ & 
$15^{\rm h}14^{\rm m}24\fs0$ & 
$+36\degr 36\arcmin 36\arcsec$ & 
95 Feb 06 -- 95 Feb 06 & $35.0$ & $0.14$ \\
\object{MS 1621.5$+$2640}  & $0.426$ & 
$16^{\rm h}23^{\rm m}36\fs0$ & 
$+26\degr 33\arcmin 36\arcsec$ & 
94 Jul 28 -- 94 Jul 30 & $43.8$ & $0.36$ \\
\object{MS 1910.5$+$6736}  & $0.246$ & 
$19^{\rm h}10^{\rm m}28\fs8$ & 
$+67\degr 41\arcmin 24\arcsec$ & 
96 Feb 22 -- 96 Mar 04 & $25.1$ & $0.60$ \\
\object{MS 2053.7$-$0449}  & $0.583$ & 
$20^{\rm h}56^{\rm m}21\fs6$ & 
$-04\degr 37\arcmin 48\arcsec$ & 
96 May 07 -- 96 May 06 & $4.9$ & $0.50$ \\
\object{MS 2137.3$-$2353}  & $0.313$ & 
$21^{\rm h}40^{\rm m}12\fs0$ & 
$-23\degr 39\arcmin 36\arcsec$ & 
92 Nov 26 -- 94 Apr 25 & $15.5$ & $0.36$ \\
\hline 
\end{tabular}
\end{table*}
%
%
\begin{table*}[t]
\caption[]{The result of standard $\beta$ model fitting.}
\label{Standard}
\begin{tabular}{lccccccc}
\hline\hline\\
Cluster name & $S_{0}$ & $\beta_{\rm fit}$ & 
${\theta_{c}}^{\star}$ & Background & $\chi^{2}$/d.o.f. & count rate\\
 & $[\:{\rm count~s^{-1}~arcsec^{-2}}\:]$ & &
 & $[\:{\rm count~s^{-1}~arcsec^{-2}}\:]$ & & $[\: {\rm count~s^{-1}}\:]$ \\
\hline\\
\object{MS 0015.9$+$1609}  &
 $4.38_{-0.48}^{+0.54}\times 10^{-6}$ &
 $0.72_{-0.08}^{+0.12}$ & 
 $42.8_{-7.4}^{+9.6}\;\arcsec$ &
 $9.88 \times 10^{-7\;\dagger}$&
 $113.34/116$ &
 0.0344 \\
 & & & $382{h_{50}}^{-1}\;{\rm kpc}$ & & $=0.98$ & ($\leq 182\arcsec$)\\
\object{MS 0302.7$+$1658}  &
 $6.68_{-2.14}^{+7.58}\times 10^{-6}$ &
 $0.62_{-0.13}^{+0.34}$ & 
 $11.3_{-6.6}^{+9.5}\;\arcsec$ &
 $1.16\times 10^{-6\;\dagger}$ &
 $77.97/84$ &
 0.00871\\
 & & & $88{h_{50}}^{-1}\;{\rm kpc}$ & & $=0.93$ & ($\leq 221\arcsec$)\\ 
\object{MS 0353.6$-$3642}  &
 $5.35_{-1.05}^{+1.27}\times 10^{-6}$ &
 $0.62_{-0.13}^{+0.28}$ & 
 $31.9_{-10.6}^{+18.8}\;\arcsec$ &
 $9.19_{-0.50}^{+0.38}\times 10^{-7}$ &
 $112.34/113$ &
 0.0449\\
 & & & $208{h_{50}}^{-1}\;{\rm kpc}$ & & $=0.99$ & ($\leq 231\arcsec$)\\
\object{MS 0451.5$+$0250}  &
 $2.94_{-0.35}^{+0.41}\times 10^{-6}$ &
 $0.74^{+0.32}_{-0.15}$ & 
 $126.0^{+55.2}_{-33.7}\;\arcsec$ &
 $1.16\times 10^{-6\;\dagger}$&
 $100.11/110$ &
 0.158\\
 & & & $587{h_{50}}^{-1}\;{\rm kpc}$ & & $=0.91$ & ($\leq 293\arcsec$)\\
\object{MS 0735.6$+$7421}  &
 $5.20_{-0.85}^{+0.95}\times 10^{-5}$ &
 $0.48_{-0.01}^{+0.02}$ & 
 $8.5_{-1.2}^{+1.5}\;\arcsec$ &
 $1.25\times 10^{-6\;\dagger}$ &
 $159.63/117$ &
 0.0945 \\
 & & & $42{h_{50}}^{-1}\;{\rm kpc}$ & & $=1.36$ & ($\leq 279\arcsec$)\\
\object{MS 1006.0$+$1202}  &
 $2.72_{-0.27}^{+0.30}\times 10^{-6}$ &
 $0.91_{-0.16}^{+0.29}$ & 
 $82.9_{-16.4}^{+24.8}\;\arcsec$ &
 $1.30\times 10^{-6\;\dagger}$ &
 $148.30/115$ &
 0.0437\\
 & & & $414{h_{50}}^{-1}\;{\rm kpc}$ & & $=1.29$ & ($\leq 155.5\arcsec$)\\
\object{MS 1008.1$-$1224}  &
 $3.91_{-0.61}^{+0.76}\times 10^{-6}$ &
 $0.63_{-0.07}^{+0.11}$ & 
 $36.4_{-8.6}^{+11.5}\;\arcsec$ &
 $1.22\times 10^{-6\;\dagger}$ &
 $166.06/116$ &
 0.0333\\
 & & & $228{h_{50}}^{-1}\;{\rm kpc}$ & & $=1.43$ & ($\leq 183\arcsec$)\\
\object{MS 1224.7$+$2007}  &
 $1.07_{-0.39}^{+0.57}\times 10^{-5}$ &
 $0.47_{-0.05}^{+0.07}$ & 
 $6.6_{-2.7}^{+4.0}\;\arcsec$ &
 $1.25\times 10^{-6\;\dagger}$ &
 $115.92/112$ &
 0.0179\\
 & & & $44{h_{50}}^{-1}\;{\rm kpc}$ & & $=1.04$ & ($\leq 170\arcsec$)\\
\object{MS 1333.3$+$1725}  & \multicolumn{5}{c}{{\it Not a cluster of galaxies}}
& 0.00367\\
\object{MS 1358.4$+$6245}  &
 $2.86_{-0.68}^{+0.83}\times 10^{-5}$ &
 $0.47_{-0.02}^{+0.02}$ & 
 $7.2_{-1.6}^{+2.2}\;\arcsec$ &
 $1.24\times 10^{-6\;\dagger}$ &
 $128.13/116$ &
 0.0484\\
 & & & $48{h_{50}}^{-1}\;{\rm kpc}$ & & $=1.10$ & ($\leq 212\arcsec$)\\
MS 1455.0$+$2232  &
 $9.60_{-1.24}^{+1.36}\times 10^{-5}$ &
 $0.64_{-0.03}^{+0.04}$ & 
 $12.3_{-1.6}^{+1.8}\;\arcsec$ &
 $1.34\times 10^{-6\;\dagger}$ &
 $154.46/117$ &
 0.102\\
 & & & $69{h_{50}}^{-1}\;{\rm kpc}$ & & $=1.32$ & ($\leq 114\arcsec$)\\
\object{MS 1512.4$+$3647}  &
 $2.01_{-0.45}^{+0.58}\times 10^{-5}$ &
 $0.59_{-0.06}^{+0.10}$ & 
 $9.4_{-2.5}^{+3.4}\;\arcsec$ &
 $1.19 \times 10^{-6\;\dagger}$ &
 $90.64/117$ &
 0.0178\\
 & & & $68{h_{50}}^{-1}\;{\rm kpc}$ & & $=0.77$ & ($\leq 129\arcsec$)\\
\object{MS 1621.5$+$2640}  &
 $3.74_{-0.58}^{+0.62}\times 10^{-6}$ &
 $117.1$ &
 $1246.8\;\arcsec$ &
 $4.04\times 10^{-6\;\dagger}$ &
 $107.36/114$ &
 $0.051$ \\
 & & & $9739{h_{50}}^{-1}\;{\rm kpc}$ & & $=0.94$ & ($\leq 166\arcsec$)\\
\object{MS 1910.5$+$6736}  &
 $3.25_{-0.81}^{+1.17}\times 10^{-6}$ &
 $0.66_{-0.12}^{+0.27}$ & 
 $29.0_{-10.3}^{+18.5}\;\arcsec$ &
 $7.20\times 10^{-7\;\dagger}$ &
 $127.30/115$ &
 0.0167\\
 & & & $157{h_{50}}^{-1}\;{\rm kpc}$ & & $=1.11$ & ($\leq 208\arcsec$)\\
\object{MS 2053.7$-$0449}  &
 $\leq 1.1\times 10^{-5}$ &
 $2/3^{\;\ddagger}$ & 
 $\cdot\cdot\cdot$ &
 $1.30 \times 10^{-6\;\dagger}$ &
 $\cdot\cdot\cdot$ &
 $\leq 0.00111$\\
\object{MS 2137.3$-$2353}  &
 $1.36_{-0.20}^{+0.23}\times 10^{-4}$ &
 $0.63_{-0.03}^{+0.04}$ & 
 $8.4_{-1.2}^{+1.4}\;\arcsec$ &
 $1.22_{-0.03}^{+0.03}\times 10^{-6}$ &
 $111.01/114$ &
 0.0728\\
 & & & $54{h_{50}}^{-1}\;{\rm kpc}$ & & $=0.97$ & ($\leq 119\arcsec$)\\
\hline\\
\end{tabular}

\noindent
$\dagger$ Fixed. See Sec.\ 2.4.\\
\noindent
$\ddagger$ Assumed to estimate the upper limit of $S_0$.\\
\end{table*}
%
%
\begin{table*}[t]
\caption[]{The result of ENF98 $\beta$ model fitting.}
\label{ENF98}
\begin{tabular}{lccccccc}
\hline\hline\\
Cluster name & $S_{0}$ & 
${\theta_{c}}^{\star}$ & Background & $\chi^{2}$/d.o.f. & count rate\\
 & $[\:{\rm count~s^{-1}~arcsec^{-2}}\:]$ & 
 & $[\:{\rm count~s^{-1}~arcsec^{-2}}\:]$ & & $[\:{\rm count~s^{-1}}\:]$ \\
\hline\\
\object{MS 0015.9$+$1609} &
 $4.21_{-0.38}^{+0.41}\times 10^{-6}$ & 
 $48.0_{-2.8}^{+3.0}\;\arcsec$ &
 $9.88\times 10^{-7\;\dagger}$ &
 $114.33/117=0.98$ &
 0.0344\\
 & & $429{h_{50}}^{-1}\;{\rm kpc}$ & & & ($\leq 182\arcsec$)\\
\object{MS 0302.7$+$1658} &
 $6.85_{-2.05}^{+2.44}\times 10^{-6}$ & 
 $15.2_{-2.8}^{+3.7}\;\arcsec$ &
 $1.15_{-0.02}^{+0.02}\times 10^{-6}$ &
 $104.45/114=0.92$ &
 0.00985 \\
 & & $119{h_{50}}^{-1}\;{\rm kpc}$ & & & ($\leq 221\arcsec$)\\
\object{MS 0353.6$-$3642} &
 $4.71_{-0.74}^{+0.82}\times 10^{-6}$ & 
 $45.6_{-4.5}^{+5.0}\;\arcsec$ &
 $9.20 \times 10^{-7}$ &
 $116.11/115=1.01$ &
 0.0449\\
 & & $297{h_{50}}^{-1}\;{\rm kpc}$ & & & ($\leq 231\arcsec$)\\
\object{MS 0451.5$+$0250}   &
 $2.90_{-0.29}^{+0.31}\times 10^{-6}$ & 
 $135.1_{-9.3}^{+10.1}\;\arcsec$ &
 $1.16\times 10^{-6\;\dagger}$ &
 $100.23/111=0.90$ &
 $0.158$\\
 & & $629{h_{50}}^{-1}\;{\rm kpc}$ & & & ($\leq 293\arcsec$)\\
\object{MS 0735.6$+$7421} &
 $1.38_{-0.17}^{+0.19}\times 10^{-5}$ & 
 $39.1_{-3.1}^{+3.3}\;\arcsec$ &
 $1.25 \times 10^{-6\;\dagger}$ &
 $297.63/118=2.52$ &
 0.0945\\
 & & $192{h_{50}}^{-1}\;{\rm kpc}$ & & & ($\leq 279\arcsec$)\\
\object{MS 1006.0$+$1202} &
 $2.83_{-0.24}^{+0.26}\times 10^{-6}$ & 
 $70.7_{-4.1}^{+4.4}\;\arcsec$ &
 $1.30 \times 10^{-6\;\dagger}$ &
 $149.78/116=1.29$ &
 0.0437 \\
 & & $353{h_{50}}^{-1}\;{\rm kpc}$ & & & ($\leq 155.5\arcsec$)\\
\object{MS 1008.1$-$1224} &
 $3.34_{-0.37}^{+0.40}\times 10^{-6}$ & 
 $52.0_{-3.9}^{+4.2}\;\arcsec$ &
 $1.22 \times 10^{-6\;\dagger}$ &
 $170.88/117=1.46$ &
 0.0333 \\
 & & $325{h_{50}}^{-1}\;{\rm kpc}$ & & & ($\leq 183\arcsec$)\\
\object{MS 1224.7$+$2007} &
 $6.69_{-2.93}^{+3.65}\times 10^{-6}$ & 
 $18.0_{-4.6}^{+8.5}\;\arcsec$ &
 $1.25 \times 10^{-6\;\dagger}$ &
 $130.19/113=1.15$ &
 0.0179 \\
 & & $119{h_{50}}^{-1}\;{\rm kpc}$ & & & ($\leq 170\arcsec$)\\
\object{MS 1333.3$+$1725} & \multicolumn{4}{c}{{\it Not a cluster of galaxies}}
 \\
\object{MS 1358.4$+$6245} &
 $8.30_{-1.38}^{+1.66}\times 10^{-6}$ & 
 $34.1_{-3.8}^{+4.2}\;\arcsec$ &
 $1.24 \times 10^{-6}$ &
 $190.23/117=1.63$ &
 0.0484\\
 & & $226{h_{50}}^{-1}\;{\rm kpc}$ & & & ($\leq 212\arcsec$)\\
MS 1455.0$+$2232 &
 $7.57_{-0.84}^{+0.91}\times 10^{-5}$ & 
 $18.2_{-1.1}^{+1.2}\;\arcsec$ &
 $1.34 \times 10^{-6\;\dagger}$ &
 $177.84/118=1.51$ &
 0.102\\
 & & $102{h_{50}}^{-1}\;{\rm kpc}$ & & & ($\leq 114\arcsec$)\\
\object{MS 1512.4$+$3647} &
 $1.60_{-0.30}^{+0.34}\times 10^{-5}$ & 
 $15.4_{-1.9}^{+1.9}\;\arcsec$ &
 $1.19 \times 10^{-6\;\dagger}$ &
 $97.89/118=0.83$ &
 0.0178\\
 & & $111{h_{50}}^{-1}\;{\rm kpc}$ & & & ($\leq 129\arcsec$)\\
\object{MS 1621.5$+$2640}$^\sharp$ &
 $4.45_{+0.79}^{-0.89}\times 10^{-6}$ &
 $63.6_{-8.0}^{+8.8}\;\arcsec$ &
 $4.04 \times 10^{-6\;\dagger}$ &
 $117.59/115=1.02$ &
 $0.051$\\
 & & $497{h_{50}}^{-1}\;{\rm kpc}$ & &  & ($\leq 166\arcsec$)\\
\object{MS 1910.5$+$6736} &
 $2.87_{-0.52}^{+0.61}\times 10^{-6}$ & 
 $38.0_{-4.8}^{+5.2}\;\arcsec$ &
 $7.20 \times 10^{-7\;\dagger}$ &
 $128.24/116=1.11$ &
 0.0167 \\
 & & $206{h_{50}}^{-1}\;{\rm kpc}$ & &  & ($\leq 208\arcsec$)\\
\object{MS 2053.7$-$0449} & 
 $\cdot\cdot\cdot$ &
 $\cdot\cdot\cdot$ &
 $1.30 \times 10^{-6}$ &
 $\cdot\cdot\cdot$ &
 $\cdot\cdot\cdot$\\
\object{MS 2137.3$-$2353} &
 $1.08_{-0.15}^{+0.16}\times 10^{-4}$ & 
 $12.7_{-0.9}^{+1.0}\;\arcsec$ &
 $1.22 \times 10^{-6\;\dagger}$ &
 $132.90/115=1.16$ &
 0.0728 \\
 & & $82{h_{50}}^{-1}\;{\rm kpc}$ & &  & ($\leq 119\arcsec$)\\
\hline\\
\end{tabular}

\noindent
$\dagger$ Fixed. See Sec.\ 2.4.\\
\end{table*}
%
%
\begin{table*}[t]
\caption[]{Flux, luminosity and temperature of the sample clusters.}
\label{FLT}
\begin{tabular}{lccccccc}
\hline\hline\\
Cluster name & \multicolumn{2}{c}{Flux ($0.1-2.4$ keV)} &
\multicolumn{2}{c}{Luminosity ($2-10$ keV)} &
\multicolumn{2}{c}{$\rm Temperature^{\dagger}$} & 
${\rm Temperature^{\ddagger}}$ \\
& \multicolumn{2}{c}{$[\:10^{-13}\;{\rm erg~s^{-1}~cm^{-2}}\:]$} & 
\multicolumn{2}{c}{$[\:{h_{50}}^{-2}\:10^{44}\;{\rm erg~s^{-1}}\:]$} &
\multicolumn{2}{c}{$[$ keV $]$} & 
$[$ keV $]$ \\
\cline{2-3} \cline{4-5} \cline{6-7}\\
& ${\rm standard^{\star}}$ & ${\rm ENF98^{\star\star}}$ &
 ${\rm standard^{\star}}$ & ${\rm ENF98^{\star\star}}$ &
 ${\rm standard^{\star}}$ & ${\rm ENF98^{\star\star}}$ & \\
\hline\\
\object{MS 0015.9$+$1609} &
 \multicolumn{2}{c}{13.0}&
 \multicolumn{2}{c}{40.7}&
 \multicolumn{2}{c}{(11.1)}&
 $8.0_{-1.0}^{+1.0\;\sharp}$\\
\object{MS 0302.7$+$1658} &
 $3.3$ & $3.8$ &
 $6.7$ & $7.7$ &
 $6.5$ & $6.8$&
 $\cdot\cdot\cdot$\\
\object{MS 0353.6$-$3642} &
 \multicolumn{2}{c}{16.2}&
 \multicolumn{2}{c}{13.4}&
 \multicolumn{2}{c}{(8.0)}&
 $8.13_{-1.7}^{+2.6\;\sharp}$\\
\object{MS 0451.5$+$0250} &
 \multicolumn{2}{c}{62.1}&
 \multicolumn{2}{c}{26.7}&
 \multicolumn{2}{c}{9.8}&
$\cdot\cdot\cdot$\\
\object{MS 0735.6$+$7421} &
 \multicolumn{2}{c}{36.1}&
 \multicolumn{2}{c}{14.8}&
 \multicolumn{2}{c}{8.2}&
$\cdot\cdot\cdot$\\
\object{MS 1006.0$+$1202} &
 \multicolumn{2}{c}{16.4}&
 \multicolumn{2}{c}{6.1}&
 \multicolumn{2}{c}{6.3}&
 $\cdot\cdot\cdot$\\
\object{MS 1008.1$-$1224} &
 \multicolumn{2}{c}{12.8}&
 \multicolumn{2}{c}{11.4}&
 \multicolumn{2}{c}{(7.6)}&
 $7.29_{-1.5}^{+2.5\;\sharp}$\\ 
\object{MS 1224.7$+$2007} &
 \multicolumn{2}{c}{6.4}&
 \multicolumn{2}{c}{4.3}&
 \multicolumn{2}{c}{(5.7)}&
 $4.3_{-0.6}^{+0.7\;\flat}$\\
\object{MS 1333.3$+$1725} &
 \multicolumn{2}{c}{$0.88^\natural$} & 
 \multicolumn{2}{c}{$0.12^\natural$} &
 $\cdot\cdot\cdot$ & $\cdot\cdot\cdot$ &
 $\cdot\cdot\cdot$ \\
\object{MS 1358.4$+$6245} &
 \multicolumn{2}{c}{17.6}&
 \multicolumn{2}{c}{13.2}&
 \multicolumn{2}{c}{(7.9)}&
 $6.50_{-0.7}^{+0.7\;\sharp}$\\ 
MS 1455.0$+$2232 &
 \multicolumn{2}{c}{37.6}&
 \multicolumn{2}{c}{17.9}&
 \multicolumn{2}{c}{(8.7)}&
 $5.45_{-0.3}^{+0.3\;\sharp}$\\ 
\object{MS 1512.4$+$3647} &
 \multicolumn{2}{c}{6.1}&
 \multicolumn{2}{c}{4.3}&
 \multicolumn{2}{c}{(5.7)}&
 $3.57_{-0.7}^{+1.3\;\sharp}$\\ 
\object{MS 1621.5$+$2640} &
 \multicolumn{2}{c}{19.5} & 
 \multicolumn{2}{c}{38.5} &
 \multicolumn{2}{c}{10.9} &
 $\cdot\cdot\cdot$ \\
\object{MS 1910.5$+$6736} &
 \multicolumn{2}{c}{6.2}&
 \multicolumn{2}{c}{2.8}&
 \multicolumn{2}{c}{5.0}&
 $\cdot\cdot\cdot$\\   
\object{MS 2053.7$-$0449} &
 \multicolumn{2}{c}{$\leq 0.597$} & 
 \multicolumn{2}{c}{$\leq 0.659$} &
 \multicolumn{2}{c}{$\leq 3.7$} &
 $\cdot\cdot\cdot$ \\
\object{MS 2137.3$-$2353} &
 \multicolumn{2}{c}{26.4}&
 \multicolumn{2}{c}{16.7}&
 \multicolumn{2}{c}{(8.5)}&
$4.37_{-0.3}^{+0.4\;\sharp}$\\ 
\hline\\
\end{tabular}
 
\noindent
$\star$ From standard $\beta$ model (See Sec.\ 2.3) fitting results.\\
\noindent
$\star\star$ From ENF98 $\beta$ model (See Sec.\ 3.3) fitting results.\\ 
\noindent
$\dagger$ From the luminosity-temperature relation
 in Arnaud \& Evrard (\cite{AE98}). See Sec.\ 2.4.\\
\noindent
$\ddagger$ From {\sl ASCA}\ observations.\\
\noindent
$\sharp$ From Mushotzky \& Scharf (\cite{Mushotzky97}).\\
\noindent
$\flat$ From Henry (\cite{Henry97}).\\
\noindent
$\natural$ Power low model with photon index of 1.8.\\
\end{table*}
%
%
\begin{table*}[t]
\caption[]{Central electron number density, cooling time, cooling radius 
 and mass-flow rate of sample clusters.}
\label{Cooling}
\begin{tabular}{lcccccccccc}
\hline\hline\\
Cluster name &
 \multicolumn{2}{c}{${{n_e}_0}$} &
 \multicolumn{2}{c}{$t_{\rm cool}$} &
 ${t_{\rm age}}$ &
 \multicolumn{2}{c}{$\theta_{\rm cool}$} &
 \multicolumn{2}{c}{$\dot{M}_{\rm cool}$} \\
 & \multicolumn{2}{c}{$[\:10^{-3}{h_{50}}^{1/2}{\rm cm^{-3}}\:]$} &
 \multicolumn{2}{c}{ $[$ Gyr $]$} &
 $[\:{h_{50}}^{-1} {\rm Gyr}\:]$ &
 \multicolumn{2}{c}{$[\: {\rm arcsec} \:]$} &
 \multicolumn{2}{c}{$[\:M_{\sun}~{\rm yr^{-1}}\:]$} \\
 \cline{2-3}\cline{4-5}\cline{7-8}\cline{9-10}\\
& ${\rm standard^{\star}}$ & ${\rm ENF98^{\star\star}}$ &
 ${\rm standard^{\star}}$ & ${\rm ENF98^{\star\star}}$ & & 
 ${\rm standard^{\star}}$ & ${\rm ENF98^{\star\star}}$ &
 ${\rm standard^{\star}}$ & ${\rm ENF98^{\star\star}}$ \\
\hline\\
\object{MS 0015.9$+$1609} &
 $7.2$ & $6.9$ & $13.9$ & $14.5$ & $11.4$ & $0$ & $0$ & $0$ & $0$ \\
\object{MS 0302.7$+$1658} &
 $18.7$ & $20.2$ & $4.8$ & $4.6$ & $12.6$ & $14$ & $18$ & $216$ & $515$ \\
\object{MS 0353.6$-$3642} &
 $7.6$ & $6.7$ & $13.3$ & $15.1$ & $13.8$ & $6$ & $0$ & $4$ & $0$ \\
\object{MS 0451.5$+$0250} &
 $3.5$ & $3.4$ & $31.6$ & $32.6$ & $15.4$ & $0$ & $0$ & $0$ & $0$ \\
\object{MS 0735.6$+$7421} &
 $40.3$ & $13.3$ & $2.5$ & $7.6$ & $15.2$ & $29$ & $35$ & $371$ & $382$ \\
\object{MS 1006.0$+$1202} &
 $4.0$ & $4.3$ & $22.2$ & $20.7$ & $15.1$ & $0$ & $0$ & $0$ & $0$ \\
\object{MS 1008.1$-$1224} &
 $6.8$ & $5.7$ & $14.0$ & $16.8$ & $14.1$ & $3$ & $0$ & $0.4$ & $0$ \\
\object{MS 1224.7$+$2007} &
 $24.6$ & $17.7$ & $3.0$ & $4.1$ & $13.7$ & $18$ & $24$ & $186$ & $397$ \\
\object{MS 1333.3$+$1725} &
 $\cdot\cdot\cdot$ & $\cdot\cdot\cdot$ &
 $\cdot\cdot\cdot$ & $\cdot\cdot\cdot$ &
 $\cdot\cdot\cdot$ &
 $\cdot\cdot\cdot$ & $\cdot\cdot\cdot$ &
 $\cdot\cdot\cdot$ & $\cdot\cdot\cdot$ \\
\object{MS 1358.4$+$6245} &
 $31.7$ & $10.4$ & $2.8$ & $8.7$ & $13.7$ & $21$ & $23$ & $348$ & $258$ \\
\object{MS 1455.0$-$2232} &
 $52.5$ & $42.2$ & $1.6$ & $2.0$ & $14.6$ & $37$ & $38$ & $1732$ & $1887$ \\
\object{MS 1512.4$+$3647} &
 $26.2$ & $21.6$ & $2.6$ & $3.1$ & $13.2$ & $22$ & $26$ & $475$ & $673$ \\
\object{MS 1621.5$+$2640} &
 $4.2$ & $5.5$ & $27.8$ & $21.2$ & $12.6$ & $0$ & $0$ & $0$ & $0$ \\
\object{MS 1910.5$+$6736} & 
 $6.3$ & $5.6$ & $12.6$ & $14.1$ & $14.8$ & $12$ & $8$ & $12$ & $3$ \\
\object{MS 2053.7$-$0449} &
 $\cdot\cdot\cdot$ & $\cdot\cdot\cdot$ &
 $\cdot\cdot\cdot$ & $\cdot\cdot\cdot$ &
 $\cdot\cdot\cdot$ &
 $\cdot\cdot\cdot$ & $\cdot\cdot\cdot$ &
 $\cdot\cdot\cdot$ & $\cdot\cdot\cdot$ \\
\object{MS 2137.3$-$2353} &
 $75.0$ & $61.3$ & $1.0$ & $1.2$ & $13.9$ & $33$ & $33$ & $2253$ & $2549$ \\
\hline\\
\end{tabular}

\noindent
$\star$ From standard $\beta$ model (See Sec.\ 2.3) fitting results.\\
\noindent
$\star\star$ From ENF98 $\beta$ model (See Sec.\ 3.3) fitting results. 
\end{table*}
%
%
\section{Cluster sample and its X-ray data}
%
%
%
\subsection{Cluster sample and its X-ray observations}
Selecting 16 distant and rich clusters with 
$L_{\rm X}(0.3-3.5 {\rm keV})>4\times10^{44}{\rm erg~s^{-1}}\:(H_{0}=50 h_{50}~
{\rm km~s^{-1}~Mpc^{-1}},\: {\Omega_{\rm m}}_0=1.0, \: {\Omega_{\Lambda}}_0 = 0.0$ and $z>0.2$)
from a complete sample of distant rich clusters selected from the EMSS,
LF94 performed a medium-deep $V, R,$ and $I$ imaging for a GLA survey
and found 6 GLAs.

X-ray observations for all the 16 clusters in the LF94 sample 
were performed by {\sl ROSAT HRI}. 
The {\sl HRI}\  data of 11 clusters were taken from 
the ROSAT Archive at Max-Planck-Institut 
f\"ur extraterrestrische Physik (MPE).
The {\sl HRI}\ data of remaining clusters were obtained by our own proposals.
The instrument {\sl HRI}\ had $\sim 5\arcsec$ FWHM resolution
and was sensitive for an energy range of $0.1-2.4$ keV
(Tr\"umper \cite{Trumper84}),
which makes {\sl HRI} the best instrument to date
to perform detailed observation of the ICM distribution 
in distant clusters.
In Table \ref{Log}, we list the log of the {\sl ROSAT HRI}\  observations.
The column density of the galactic hydrogen in 6th column in the Table \ref{Log} is calculated
using {\sl EXSAS}
\footnote{{\sl Extended Scientific Analysis System} \/ 
provided by ROSAT Science Data Center at MPE}
command `CALCULATE/GALACTIC\_NH' which calculates the galactic hydrogen column density 
toward the specified direction and is based on Dickey \& Lockman (\cite{DL90}).
Our observation revealed that \object{MS 1333.3$+$1725} was {\it not} a cluster 
but an X-ray point source.
%
%
\subsection{ROSAT HRI data reduction}
We used {\sl EXSAS}
and {\sl XSPEC}\ \footnote{provided by NASA/Goddard Space Flight Center.}
analysis packages to reduce the data. 
The position of point sources higher than $3 \sigma$
is determined via standard source detection pipeline for {\sl HRI}\ data 
in {\sl EXSAS}. 
The cluster center was determined as the brightest X-ray peak.
Accuracy of {\sl ROSAT HRI}\  pointing was checked with both 
Hambrug RASS Catalog of Optical ID (HRASSCAT)
and ROSAT SIMBAD identifications (ROSID)\footnote{
both are available via High Energy Astrophysics Science Archive 
Research Center (HEASARC) Online Service, 
provided by NASA/Goddard Space Flight Center. 
{\sf http://heasarc.gsfc.nasa.gov/docs/frames/mb\_w3browse.html}}. 
Positions of sources higher than $3 \sigma$ were compared with positions of 
objects cataloged in HRASSCAT or ROSID.
The vignetting of the {\sl ROSAT HRI}\  was less than 5\% 
within a radius of 5 arcminutes (i.e.\ 600 pixels)
from the detector center 
at all energy range of $0.1-2.4$ keV for which the {\sl ROSAT HRI}\  was sensitive.
Thus we restricted our analysis to the inner 600 pixels of each image, 
where the background can be regarded as flat.
%
%
\subsection{Data analyses}
Photon event tables were binned into radial rings
to make azimuthally averaged surface brightness profiles. 
The width of each ring was determined in order that 
the number of photons in each bin become greater than or equal to 
25 to ensure that $\chi^2$ fitting could be performed, 
and that the size of each bin become greater than or equal to 
$2.5\arcsec \sim {\rm FWHM}/2$.
A radial surface brightness profile was then constructed by
summing up the counts in each bin. 
Note that all the contaminating point sources 
which were higher than $3 \sigma$ were excluded from the photon counting. 

The radial surface brightness profile was fitted via $\chi^2$-minimization 
routine to (Cavaliere and Fusco-Femiano \cite{Cavaliere76})
\begin{equation}
S(\theta) = S_{0}\left[1+\left(\frac{\theta}{\theta_c}\right)^2
\right]^{-3\beta_{\rm fit}+1/2}+B,
\label{sbm}
\end{equation}
where $S_0$ is the central surface brightness, 
$\theta_c$ is the angular core radius, 
and $B$ is the background. 
It was physically interpreted that $\beta_{\rm fit}$ described 
the ratio of the kinetic energy per unit mass of the member galaxy to that in the ICM
if cluster galaxy and the ICM distributions are isothermal and 
galaxy velocity dispersion is isotropic. 
Therefore, the surface brightness distribution described by the Eq.\ (1) was called 
isothermal $\beta$ model. 
However, it seems that such a situation is far from the reality
(Lubin \& Bahcall \cite{Lubin93}; Bahcall \& Lubin \cite{Bahcall94}).
Therefore, Eq.\ (1) has no meaning more than a conventional fitting model.
We call the model $\beta$ described by the Eq.\ (1) `{\sl standard $\beta$ model}' in this paper.
The background value was first determined via the above fitting.
Removing the background from the image we then checked  
that the radial source photon counts remained constant
outside the cluster source region, if not we modified the background
value accordingly, and we re-did the standard $\beta$ model fitting.
In Table \ref{Standard}, 
we list the standard $\beta$ model fitting result.
The bracketed numbers in 6th column is the edge within which
source photon numbers are counted.
%
%
\subsection{X-ray properties of the sample clusters}
We list fluxes, luminosities and temperatures
of the sample clusters calculated using the best-fit values of
the standard $\beta$ model fitting on the left side of each column
in Table \ref{Standard}.
Fluxes in the $0.1-2.4$ keV band were calculated
on the {\sl XSPEC} using Raymond-Smith thermal plasma model 
assuming 30\% of solar metallicity.
The calculation of the flux requires the value of the temperature. 
For clusters who have no {\sl ASCA} observation, 
the temperature of 6 keV was first assumed 
and the temperature was calculated iteratively 
using the 2--10 keV X-ray luminosity$-$temperature ($L_{\rm X}-T$) relation
of Arnaud \& Evrard (\cite{AE98}, henceforth AE98)
until the temperature converges.
When {\sl ASCA} temperature ($T_{ASCA}$) was available, we used it
to compute the X-ray flux and luminosity of the clusters.
This luminosity was then used to compute the
expected temperature from the $L_{\rm X}-T$ relation
(we specify this in Table \ref{FLT} by bracketing the temperatures
with parentheses in column 6).

Although the $L_{\rm X}-T$ relation of AE98
was derived from nearby cluster sample,
we used it for our distant cluster sample 
assuming no evolution of $L_{\rm X}-T$ relation.
David et al. (\cite{David93}, henceforth D93) also derived 2--10 keV 
$L_{\rm X}-T$ relation in (${\Omega_{\rm m}}_0$,${\Omega_{\Lambda}}_0$)=($0$, $0$) cosmology
using Raymond-Smith thermal plasma model 
but they assumed 50 \% of solar metallicity 
which is too high value for the ICM.
Since AE98 showed the $L_{\rm X}-T$ relation in
(${\Omega_{\rm m}}_0$,${\Omega_{\Lambda}}_0$)=($1$, $0$) cosmology,
we re-plotted the $L_{\rm X}-T$ diagram in
(${\Omega_{\rm m}}_0$,${\Omega_{\Lambda}}_0$)=($0.3$, $0.7$) cosmology.
The $L_{\rm X}-T$ relation we used is 
$T_{\rm X}  = 10^a{L_{\rm X}}^b$, where
$T_{\rm X}$ is X-ray temperature in keV and 
${L_{\rm X}}$ is 2--10 keV X-ray luminosity in ${\rm erg~s^{-1}}$,
and where $(a,b)=(-12.47\pm 0.72, 0.296\pm 0.016)$.
As AE98 discussed, 
their $L_{\rm X}-T$ relation has the slope of $1/2.88=0.347$
steeper than D93
but the $L_{\rm X}-T$ relation of AE98 
in (${\Omega_{\rm m}}_0$,${\Omega_{\Lambda}}_0$)=($0.3$, $0.7$) cosmology
is rather consistent with D93.
Mushotzky \& Scharf (\cite{Mushotzky97}) reported
the $L_{\rm X}-T$ relation for distant clusters
and showed that there was no evolution of 
$L_{\rm X}-T$ relation up to $z=0.5$ 
comparing with the sample of D93.
In Table \ref{Cooling}, 
we list central electron number densities (${{n_e}_0}$), 
central cooling times ($t_{\rm cool}$),
ages of the universe at the cluster's redshift (${t_{\rm age}}$),
cooling radii ($\theta_{\rm cool}$),
and mass-flow rates ($\dot{M}_{\rm cool}$) for sample clusters.
The central electron number density was calculated using rest frame
0.5--2.0 keV {\sl HRI} luminosity assuming the gaunt factor of
$0.9[h\nu/(k_{B}T_{\rm X})]^{-0.3}$ (Henry \& Henriksen \cite{HH86}),
where $h$ is the Plank constant, $\nu$ is the frequency,
$k_{B}$ is the Boltzmann constant, and $T_{\rm X}$ is the X-ray temperature,
and one-ninth of the hydrogen number density for the helium number density is assumed.
The cooling radius is defined as the radius where $t_{\rm cool} = t_{\rm age}$.
The cooling mass-flow rate was calculated using Eq.~(2) in Fabian
(\cite{Fabian94}).
%
%
\section{Cluster mass models and their lens properties}

In this section,
we summarize
lens properties of three basic cluster mass distribution models.

\subsection{The singular isothermal sphere (SIS) model}
Although we do not use the SIS model,
the SIS model is employed by HF97 and then 
we give its lens properties to compare with those of lens models we employed.
The density profile of a cluster described by the SIS model is
\begin{equation}
\rho_{\rm SIS}(r) = \frac{\sigma^2}{2\pi G}\frac{1}{r^2} ,
\end{equation}
where $\sigma^2$ is the velocity dispersion of the cluster.
The lens equation of the SIS model thereby becomes
\begin{equation}
y = x - \frac{x}{|x|},
\end{equation}
where $y$ [$x$] is the angle between a source [an image] and the lens center 
in the unit of the Einstein ring radius of SIS model $\theta_{\rm E}^{SIS}$:
\begin{equation}
\theta_{\rm E}^{SIS} = 4 \pi \frac{\sigma^2}{c^2} \frac{D_{LS}}{D_{OS}},
\end{equation}
where $D_{LS}$ [$D_{OS}$] is the angular diameter distance 
from the lens [the observer] to the source.

If we assume all background sources circular,
length-to-width ratio of a GLA equals to
the ratio of the tangential stretching rate to the radial stretching rate or vise versa.
(As one can see in the lens equation, 
the radial stretching rate of the SIS model is always unity.)
The moduli of the eigenvalues of the Jacobian matrix of the lens mapping are
\begin{eqnarray}
\lambda_{\rm t}({\rm tangential})
&\equiv& \left|\frac{\theta_{S}}{\theta_{I}}\right|
= \left|1 - \frac{1}{x} \right|
= \left|1 - \frac{\theta_{E}}{\theta_{I}} \right|  ,\\
\lambda_{\rm r}({\rm radial})
&\equiv& \left|\frac{d \theta_{S}}{d \theta_{I}}\right|
= 1 .
\end{eqnarray}
and the length-to-width ratio of an arc is described as
\begin{equation}
R \equiv \frac{L}{W} = \frac{\lambda_{r}}{\lambda_{t}} =
\left| 1 - \frac{\theta_{E}}{\theta_{I}} \right|^{-1} .
\end{equation}

Following Wu \& Hammer (\cite{Wu93}),
a giant arc is defined as the image which has
\begin{equation}
R \geq \epsilon_{\rm th} \equiv 10,
\end{equation}
where $\epsilon_{\rm th}$ is the threshold value of length-to-width ratio for
identifying a giant arc (Paper I).
The position of arc to be giant has to fulfill the following inequalities:
\begin{equation}
0 < |\theta_{I}| \leq
max(\frac{\epsilon_{\rm th}}{\epsilon_{\rm th}-1}\theta_{E},
\frac{\epsilon_{\rm th}}{\epsilon_{\rm th}+1}\theta_{E}) =
\frac{\epsilon_{\rm th}}{\epsilon_{\rm th}-1}\theta_{E}.
\end{equation}
The cross section in the source plane for forming arcs with $R \geq \epsilon_{\rm th}$ hence reads
\begin{equation}
\sigma_{G}^{\rm SIS} = \pi \left( \frac{D_{OS}\theta_{E}}{\epsilon_{\rm th}-1} \right)^{2}.
\end{equation}
%
%
\subsection{The isothermal $\beta$ model}
Under the assumptions of spherical symmetry, 
isothermality, and the hydrostatic equilibrium for the ICM distribution,
we can calculate the total gravitational mass within a radius $r$
of a cluster whose ICM spatial distribution is described
with the standard $\beta$ model (Eq.~\ref{sbm}): 
\begin{equation}
M_{\rm cl}^{\beta}(\leq r) = 
3 \beta \frac{k_{\rm B}T_{\rm X}}{G \mu m_{p}/r_c}
\frac{(r/r_{c})^{3}}{1+(r/r_{c})^{2}},
\end{equation}
where $G$ is the gravitational constant,
$\mu$ is the mean molecular weight,
$m_{p}$ is the proton mass,
and $r_{c}$ is a physical core radius; $r_{c} = D_{OL}\theta_{c},$ where 
$D_{OL}$ is the angular diameter distance from the observer to the lens. 
We employ $\mu = 13/21$.
Henceforth we call this model `{\it isothermal $\beta$ model}'.

The density profile of the isothermal $\beta$ model is given by
\begin{equation}
\rho(r) = \rho_{0}\frac{(r/r_{c})^{2}+3}{[1+(r/r_{c})^{2}]^{2}};\;\;\;\;\;\;
\rho_{0} \equiv
\frac{3\beta}{4\pi}\frac{k_{\rm B}T_{\rm X}}{G \mu m_{p}/r_{c}}
\frac{1}{r_{c}^{3}} .
\end{equation}

The lens equation of the isothermal $\beta$ model becomes  
\begin{equation}
y = x - {\cal D}\frac{x}{[x^{2}+1]^{1/2}},
\end{equation}
where $y$ [$x$] is the angle between the lens center and the source [the image] 
in the unit of $\theta_{c}$, and ${\cal D}$ is called the lens parameter:
\begin{equation}
{\cal D} \equiv
\frac{6\pi\beta}{\theta_{c}}\frac{k_{\rm B}T_{\rm X}}{\mu m_{p}c^{2}}
\frac{D_{LS}}{D_{OS}} .
\end{equation}
When ${\cal D} \leq 1$, 
the lens always produces a single image of a source. 
On the other hand, 
the lens is able to form three images of a source when ${\cal D} > 1$.

The moduli of the eigenvalues of the Jacobian matrix of the lens mapping are
\begin{eqnarray}
\lambda_{\rm t}({\rm tangential})
&=& \left|1 - \frac{\cal D}{[x^{2}+1]^{1/2}} \right| ,\\ 
\lambda_{\rm r}({\rm radial})
&=& \left|1 - \frac{\cal D}{[x^{2}+1]^{3/2}} \right| ,
\end{eqnarray}
and the length-to-width ratio of an arc is
\begin{equation}
R_{\rm t}({\rm tangential\;arc}) \equiv
\frac{\lambda_{\rm r}}{\lambda_{\rm t}} =
\left|\frac{1 - {\cal D} [x^{2}+1]^{-3/2}}{1 - 
{\cal D} [x^{2}+1]^{-1/2}}\right| ,
\end{equation}
or
\begin{equation}
R_{\rm r}({\rm radial\;arc}) \equiv
\frac{\lambda_{\rm t}}{\lambda_{\rm r}} =
\left|\frac{1 - {\cal D} [x^{2}+1]^{-1/2}}
{1 - {\cal D} [x^{2}+1]^{-3/2}}\right| .
\end{equation}
%
%
\begin{table*}
\caption[]{Parameters of ENF98-NFW (and MSS98-${\rm NFW^{\star}}$) model.}
\label{ENFNFW}
\begin{tabular}{lccccccccc}
\hline\hline\\
Cluster name & \multicolumn{2}{c}{$\delta_{c}$} & \multicolumn{2}{c}{$r_{s}$} &
\multicolumn{2}{c}{$r_{\rm vir}$} & \multicolumn{2}{c}{$M_{\rm vir}$} & 
$T_{\rm vir}$ \\
 & \multicolumn{2}{c}{$[\:\rho_{\rm crit}{(z)}\:]$} & 
\multicolumn{2}{c}{$[\:{h_{50}}^{-1}{\rm Mpc}\:]$} & 
\multicolumn{2}{c}{$[\:{h_{50}}^{-1}{\rm Mpc}\:]$} & 
\multicolumn{2}{c}{$[\: 10^{14} {h_{50}}^{-1} M_{\sun} \:]$} & 
$[$ keV $]$ \\ 
\cline{2-3} \cline{4-5} \cline{6-7} \cline{8-9}\\
& ENF98 & ${\rm MSS98^{\star}}$ & ENF98 & ${\rm MSS98^{\star}}$ & 
ENF98 & ${\rm MSS98^{\star}}$ & ENF98 & ${\rm MSS98^{\star}}$ & ENF98 \\
\hline\\
\object{MS 0015.9$+$1609} &
6407 &  1041 & 1.14 & 1.74  & 5.78 & 3.77 & 145 & 35 & 34.9 \\
\object{MS 0302.7$+$1658} &
21606 &  9934 & 0.315 & 0.400 & 2.73 & 2.50 &  13 & 8.9 & 6.5 \\
\object{MS 0353.6$-$3642} &
8094 & 2514 & 0.787 & 0.945 & 4.60 & 3.28 &  51 & 16 & 15.5 \\
\object{MS 0451.5$+$0250} &
4676 &  516 & 1.67 & 2.67  &  7.92 & 4.40 &  214 & 37 & 37.6 \\
\object{MS 0735.6$+$7421} &
18019 & 53857 & 0.509 & 0.191 &4.28 & 1.61 &35 &  5.5 & 11.3 \\
\object{MS 1006.0$+$1202} &
5795 &  804 & 0.935 & 1.88 &4.85 & 3.88 & 51 & 28 & 14.5 \\
\object{MS 1008.1$-$1224} & 
7035 & 1148 & 0.861 & 1.04 & 4.76 & 2.49 &  55 & 8.1 & 16.0 \\
\object{MS 1224.7$+$2007} & 
60568 & 33733 & 0.315 & 0.200 &4.23 & 2.12 & 14 & 5.1 &  4.6 \\
\object{MS 1333.3$+$1725} & 
$\cdot\cdot\cdot$ & $\cdot\cdot\cdot$ & $\cdot\cdot\cdot$ &
$\cdot\cdot\cdot$ & $\cdot\cdot\cdot$ & $\cdot\cdot\cdot$ &
 $\cdot\cdot\cdot$ & $\cdot\cdot\cdot$ & $\cdot\cdot\cdot$\\
\object{MS 1358.4$+$6245} & 
12451 & 23779 & 0.599 & 0.218 &4.20 & 2.00 & 40 & 3.9 & 13.1 \\
\object{MS 1455.0$+$2232} & 
54564 & 11405 &0.270 & 0.314 &3.53 & 2.13 & 21 &  5.2 &  8.2 \\
\object{MS 1512.4$+$3647} &  
24594 & 9266 &0.294 & 0.309 &2.72 & 1.89 & 11 & 4.1 &  5.9 \\
\object{MS 1621.5$+$2640} & 
$\cdot\cdot\cdot$ & $\cdot\cdot\cdot$ & $\cdot\cdot\cdot$  & 
$\cdot\cdot\cdot$ &$\cdot\cdot\cdot$ & $\cdot\cdot\cdot$ & 
$\cdot\cdot\cdot$  & $\cdot\cdot\cdot$ & $\cdot\cdot\cdot$\\
\object{MS 1910.5$+$6736} &  
7345 & 3146 &0.546 & 0.714 &3.12 & 2.80 & 14 & 10 &  6.3 \\
\object{MS 2053.7$-$0449} & 
$\cdot\cdot\cdot$ & $\cdot\cdot\cdot$ & $\cdot\cdot\cdot$  & 
$\cdot\cdot\cdot$ & $\cdot\cdot\cdot$ & $\cdot\cdot\cdot$  & 
$\cdot\cdot\cdot$ & $\cdot\cdot\cdot$ & $\cdot\cdot\cdot$\\
\object{MS 2137.3$-$2353} & 
74641 & 20589 & 0.217 & 0.245 & 3.18 & 2.13 & 17 &  5.0 &  7.3 \\
\hline\\
\end{tabular}

\noindent
$\star$ See Sec.\ 5.

\end{table*}
%
%
\subsection{The universal dark matter halo profile model}
NFW made series of studies on the  
density profiles of dark matter halos formed through 
a gravitational collapse in hierarchically clustering cold dark matter (CDM) cosmology
using N-body simulations. 
They concluded that 
dark matter halo reached a density profiles with a universal shape 
that did not depend on their mass ranging 
from dwarf galaxy halos to those of rich clusters, 
nor on the power spectrum of initial fluctuations, 
nor on the cosmological parameters
through dissipationless hierarchical clustering.
They found that the universal density profile can be specified by giving 
two parameters: the halo mass and the halo characteristic (dimensionless) 
density $\delta_c$:
\begin{equation}
\frac{\rho_{\rm NFW}(r)}{\rho_{\rm crit}} = 
\frac{\delta_c}{(r/r_s)[1+(r/r_s)]^2},
\end{equation}
where $r_s$ is a scale radius related to the virial mass of the halo, 
and $\rho_{\rm crit}$ is the critical density:
\begin{equation}
\rho_{\rm crit}(z) = \frac{3H(z)^2}{8\pi G} = 
\frac{3{H_0}^2}{8\pi G}\left[ {\Omega_{\rm m}}_0(1+z)^3+{\Omega_{\Lambda}}_0\right],
\end{equation}
for ${\Omega_{\rm m}}_0+{\Omega_{\Lambda}}_0=1$ cosmology.

The universal dark halo profile becomes
one parameter function in the spherical top-hat model
(e.g.\  ENF98). 
The virial mass $M_{\rm vir}$ is defined to be the mass contained within 
the radius, $r_{\rm vir}$, that enclose a density contrast 
$\Delta_c:M_{\rm vir}= (4\Delta_c\pi /3)\rho_{\rm crit}{r_{\rm vir}}^3$. 
This density contrast depends on the value of $\Omega$ and
can be approximated by 
\begin{equation}
\Delta_c(\Omega, z) = 178 \;\Omega(z)^{0.45},
\end{equation}
where 
\begin{equation}
\Omega(z) = \frac{{\Omega_{\rm m}}_0(1+z)^3}{{\Omega_{\rm m}}_0(1+z)^3+{\Omega_{\Lambda}}_0},
\end{equation}
for ${\Omega_{\rm m}}_0+{\Omega_{\Lambda}}_0=1$ cosmology.
This definition shows that the ratio of virial radius to scale radius,
which is denoted by the `concentration' parameter: ${\cal C}=r_{\rm vir}/r_{s}$, 
uniquely related to 
$\delta_c$ by
\begin{equation}
\delta_c = \frac{\Delta_c}{3}\frac{{\cal C}^3}{\ln (1+{\cal C}) - 
{\cal C}/(1+{\cal C})}.
\end{equation}
The structure of a halo mass $M_{\rm vir}$ is hence completely specified 
by a single parameter.
The lens equation of the universal dark matter halo profile model
is described as
(Bartelmann \cite{B96})
\begin{equation}
y = x - {\cal K}\frac{f(x)}{x},
\end{equation}
where $y$ ($x$) is the angle between the lens center and a source (an image) 
in the unit of the angular scale radius ($r_s/D_{OL}$),
${\cal K}$ is a constant coefficient:
\begin{equation}
{\cal K} \equiv 16 \pi \frac{G}{c^2}\rho_{\rm crit}\delta_c r_s 
\frac{D_{OL}D_{LS}}{D_{OS}},
\end{equation}
and 
\begin{equation}
f(x) = \left\{
\begin{array}{ll}
\ln\frac{x}{2}+\frac{2}{\sqrt{1-x^2}}{\rm arctanh}
\sqrt{\frac{1-x}{1+x}}\;\;\;\;\;(x<1) \\ \\
1-\ln 2\;\;\;\;\;(x=1)\\ \\
\ln\frac{x}{2}+\frac{2}{\sqrt{x^2-1}}\arctan
\sqrt{\frac{x-1}{x+1}};\;\;\;\;(x>1).
\end{array}\right.
\end{equation}

ENF98 performed 
cosmological hydrodynamical and particle simulations 
to examine the evolution of X-ray emitting hot gas in clusters
in a flat (${\Omega_{\rm m}}_0+{\Omega_{\Lambda}}_0=1$), 
low-density (${\Omega_{\rm m}}_0 =0.3$) CDM cosmology. 
They showed that radial density profiles of gas in relaxed
clusters are well described by the standard $\beta$ model.
Table 3 in ENF98 enables us to calculate $\delta_c$ and $r_s$ from the X-ray data.
They have shown that the $\beta$ values of most of their simulated  
clusters are around 0.79 
if we exclude the cluster which is likely in dynamically non-equilibrium state. 
Parameters of the universal density profile is thereby 
specified with the observed core radius $r_c$
and the normalized central gas densities
$\delta_{\rm gas}$ using the relations of 
$$r_s = 2.65 r_c$$ and $$\delta_c = 3.99 \delta_{\rm gas}$$ 
found in their result for $\beta_{\rm fit} = 0.79$. 
In applying this model to our data,
we had to remind that the best-fit $\beta$ value got artificially lower as 
the central surface brightness got closer to the background surface brightness
(e.g.\ Bartelmann \& Steinmetz \cite{Bartelmann96}). 
In general, 
the central surface brightnesses of high redshift clusters are very low and close to 
the background of {\sl ROSAT HRI}.
It is hence likely that the best-fit values of $\beta$ shown
in Table \ref{Standard} are biased by this effect.  
To overcome this problem,
we employed another $\beta$ model (henceforth `{\sl ENF98 $\beta$ model}')
fitting in which $\beta$ was fixed to  
the median value of $\beta$ in Table 3 of ENF98;
$<\beta_{\rm fit}>_{\rm median} = 0.79$. 
In this procedure,
we implicitly assumed that if the radial profile was resolved up to 
the virial radius, the $\beta$ should be $\sim 0.79$.
We list the ENF98 $\beta$ model fitting result in Table \ref{ENF98}.
We also list fluxes, luminosities and temperatures
of the sample clusters calculated using the best-fit values of
the ENF98 $\beta$ model fitting on the right hand side of each
column in Table \ref{FLT}.
We also list central electron number densities, central cooling times,
ages of the universe at the redshift of each cluster, cooling radii,
and mass-flow rates ($\dot{M}_{\rm cool}$) for the sample clusters
on the right hand side in each column in Table \ref{Cooling}.
On the left hand side of each column in Table \ref{ENFNFW}, 
we list values of parameters of the NFW model
(henceforth {\it ENF98-NFW model}\ )
derived from {\sl ROSAT HRI}\  data 
using the procedure described above.
The virial mass was evaluated using the equation 
(Makino et al.\ \cite{Makino98})
\begin{equation}
M_{\rm vir} = 4 \pi \delta_c \rho_{\rm crit} {r_s}^3
\left[ \ln(1+{\cal C})+\frac{\cal C}{1+{\cal C}}\right] .
\end{equation}
For the ENF98-NFW model, one can evaluate its virial temperature $T_{\rm vir}$:
\begin{equation}
k_{\rm B}T_{\rm vir} = \frac{1}{2}\mu m_p\frac{GM_{\rm vir}}{r_{\rm vir}}
\end{equation}
using {\it only} the information of ICM spatial distribution; 
$\beta$ and $\theta_c$. 
Some clusters have virial temperatures 
which are much higher than temperatures obtained by {\sl ASCA} or
obtained using the $L_{\rm X}-T$ relation of AE98. 
%
%
\begin{figure}
 \resizebox{\hsize}{!}{\rotatebox{270}{\includegraphics{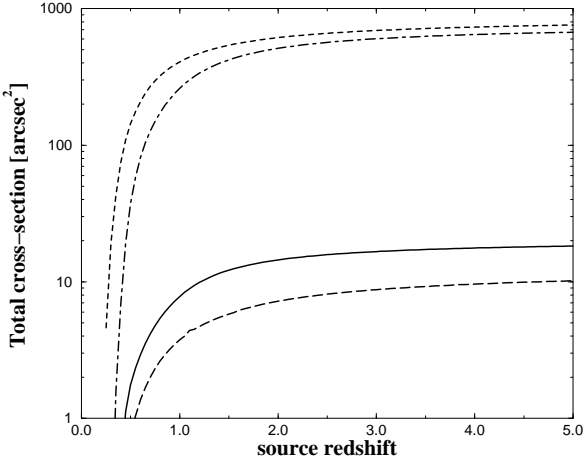}}}
 \caption{Total cross-section against source redshift.
 The total cross-section is calculated by summing up all the cross-section
 of each sample cluster.
 A cross-section for a sample cluster 
 to make giant arcs of background galaxies with a lens model
 is an area on the source plane in which 
 background circular galaxies show their images with length-to-width ratio 
 greater than or equal to 10 in the image plane.
 Clusters \object{MS 1333.3+1725},\object{MS 1621.5+2640}, and 
 \object{MS 2053.7-0449} are excluded in the calculation 
 with isothermal $\beta$\ model, ENF98-NFW model, 
 and MSS98-NFW model (See Sec.\ 5).
 Dashed line: the SIS model.
 Solid line: the isothermal $\beta$ model.
 Dot-dashed line: the ENF98-NFW model.
 Long-dashed line: the MSS98-NFW model (See Sec.\ 5).}
 \label{CS}
\end{figure}
%
%
\subsection{Total cross-sections to make giant arcs}
In Fig.~\ref{CS}, 
we show total cross-sections of LF94 sample 
to make giant arcs of background galaxies 
assuming that the background source galaxies are circular.
Dashed line, solid line, dot-dashed line, and  long-dashed line
respectively represent
the total-cross section to make giant arcs 
calculated with the SIS model, the isothermal $\beta$ model,
the ENF98-NFW model, and
the MSS98-NFW model (we discuss this model in Sec.\ 5).
A cross-section of a sample cluster 
to make giant arcs of background galaxies with a lens model
is an area on the source plane in which 
background galaxies show their images with length-to-width ratio greater than
or equal to 10 in the image plane.
The total cross-section is calculated by summing up all the cross-sections
of sample clusters.
We calculated the total cross-section with the SIS model
using $\sigma-T$ relation used in HF97.
Note that clusters \object{MS 1333.3+1725},\object{MS 1621.5+2640}, and 
\object{MS 2053.7-0449} are excluded in the calculation 
with the isothermal $\beta$\ model, the ENF98-NFW model,
and MSS98-NFW model (See Sec.\ 4.3).
As shown in Fig.~\ref{CS},
the total cross-sections calculated with the ENF98-NFW model and 
that with the SIS model are comparable for all the source redshifts.
Considerably larger total cross-section of the ENF98-NFW model
than that of the isothermal $\beta$ model is 
due to their much higher $T_{\rm vir}$.
The main reason why HF97 could reproduce the observed number of GLAs
in the LF94 sample is considerably high temperatures
which they overestimated with inappropriate $L_{\rm X}-\sigma$ relation.
%
%
\subsection{Properties of the giant luminous arcs}
In Table \ref{distance}, 
observational properties of GLAs found in the LF94 sample are
summarized as follows; major axis length:  $l$, length-to-width ratio: $l/w$,
distance from the cluster center: $d$, 
apparent magnitude: $m$,
the projected mass within Einstein ring radii $\theta_E$
regarding $d=\theta_E$:
\[
 M_{GLA}^{Einstein}(\theta_E)=
 \frac{c^2}{4\pi G}\frac{D_{OS}}{D_{OL}D_{LS}}\pi({D_{OL}\theta_E})^2,
\]
the projected mass deduced from the isothermal $\beta$ model:
\[
 M_{\rm X}^{IT\beta}(\theta_E)=
 \frac{3\pi\beta}{2}\frac{k_{B}T_{\rm X}}{G \mu m_{p}/r_{c}} 
\frac{(\theta_E/\theta_{c})^{2}}{[1+(\theta_E/\theta_{c})^{2}]^{1/2}},
\]
and the projected mass deduced from the ENF98-NFW model:
\[
 M_{\rm X}^{\rm NFW}(\theta_E)=4\pi \rho_{\rm cirt}\delta_c {r_s}^3 f(x_E),
\]
where $x_E=D_{OL}\theta_E/r_s$, assuming source redshift $z_s=1$.
%
%
%
%
%
\section{The statistics of giant luminous arcs}
In this section,
we examine 
the expected numbers of GLAs 
predicted by the isothermal $\beta$ model and the ENF98-NFW model
using the numerical code described in detail in Paper I.
%
%
\subsection{The theoretical model of galaxy evolution}
As is in Paper I, 
prescriptions of properties of background galaxies are essentially
the same as those in Yoshii \& Takahara (\cite{Yoshii88}) and 
Yoshii \& Peterson (\cite{Yoshii91}).
We adopted the galaxy type mixing ratio of (E/S0, Sab, Sbc, Scd, Sdm)=
(0.215, 0.185, 0.160, 0.275, 0.165) given by Pence (\cite{Pence76}).
The $K$ and $E$ corrections for each type were calculated 
using the type-dependent,
present day spectral energy distribution (SED)
updated by Yoshii \& Peterson (\cite{Yoshii91})
and the type-dependent galaxy luminosity evolution models 
by Arimoto \& Yoshii (\cite{Arimoto86} \& \cite{Arimoto87}),
except for the UV light of E/S0 galaxies.
We adopted the UV-intermediate NGC3379 SED (henceforth `case I') and 
the UV-bright NGC4649 SED (henceforth `case II') as the SED for E/S0 galaxies.
These models are reliable, especially at low redshift of $z<1$ 
(Yoshii \& Takahara \cite{Yoshii88}, Totani et al.\ \cite{Totani97}).
Although recent observations of galaxies and their evolution at $z>1$ 
(e.g.\ Roche et al.\ \cite{Roche98} and references therein)
have considerably been improving our knowledge of them,
these models are still in good agreement with such recent observations;
e.g.\ data of star formation history taken by the Hubble Space Telescope 
(Madau et al.\ \cite{Madau96}) or 
galaxy number counts (Yoshii \& Peterson (\cite{Yoshii91}).
We discuss some effects of the recent new knowledge for the galaxy evolution
(e.g.\ a major merger of galaxies) on our GLA statistics bellow. 
The luminosity function of all galaxy types was assumed to be same and 
was taken from Efstathiou, Ellis \& Peterson (\cite{Efstathiou88}), 
namely the Schechter function with $\alpha=-1.07$,
$\phi_*=1.56\times 10^{-2}h^3{\rm Mpc^{-3}}$ and 
$M^*_B=-19.39+5{\rm log}h$ 
($h$ is the Hubble constant in the unit of $100 {\rm km~s^{-1}~Mpc^{-1}}$).
The galaxy formation epoch was assumed to be $z_F=5$.
The absolute magnitude in $B$ band for each type of galaxies
was converted into $V$ band magnitude using the relation $M_V=M_B+(V-B)$ 
where $(V-B)=-1.03$ for ${\rm E/S0}$, 
$-0.79$ for Sab, $-0.64$ for Sbc, $-0.56$ for Scd and $-0.46$ for Sdm.
No evolution in the comoving galaxy number density
was assumed for the galaxy luminosity function.
The luminosity profiles and the effective radii for each galaxies were 
modeled as the same way in Paper I.
The intrinsic shape of the source galaxy image was assumed to be circular.
%
%
\subsection{The detection condition and the arc identification scheme}
The detection condition in the LF94 arc survey were taken into account
as the same way in Paper I.
The adopted GLA identification scheme was also same as described in 
section 2.4 of Paper I.
The length-to-width condition and the apparent magnitude condition 
were applied for the images smeared by the seeing and
by the limiting surface brightness.
The threshold value for the axis ratio and the apparent V magnitude respectively
were set to be $\epsilon_{\rm th}=10$ and $m_V({\rm arc})=22.5$. 
%
%
\begin{table*}
\caption[]{Properties of the giant luminous arcs from TABLE 1 in LF94.}
\label{distance}
\begin{tabular}{lccccccc}
\hline\hline\\
Cluster name & $l$ & $l/w$ & $d$ &
 $m$ & $M_{\rm GLA}^{Einstein}$ & $M_{\rm X}^{IT\beta}$
 & $M_{\rm X}^{\rm NFW}$ \\
 & $[\: {\rm arcsec} \:]$  & & $[\: {\rm arcsec} \:]$ &
 $[\: {\rm magnitude} \:]$ & $[\: 10^{14} M_{\sun} \:]$ &
 $[\: 10^{14} M_{\sun} \:]$ & $[\: 10^{14} M_{\sun} \:]$\\
\hline\\
\object{MS 1006.0$+$1202} & $5.9$ & $10.5$ & $26$ & $V=21.92$ &
 $1.2$ & $0.37$ & $7.3$\\
\object{MS 1008.1$-$1224} & $4.5$ & $10$   & $47$ & $V=21.53$ &
 $5.5$ & $1.8$ & $4.3$\\
\object{MS 1621.5$+$2640} & $9.8$ & $19.6$ & $16^{\;\star}$ & $V=21.16$ &
 $1.0$ & $3.4$ & $\cdot\cdot\cdot$\\
\object{MS 1910.5$+$6736} & $6.1$ & $10.5$ & $67$ & $V=22.29$ &
 $8.7$ & $1.8$ & $4.0$\\
\object{MS 2053.7$-$0449} & $10.5$ & $17.5$ & $15.8$ & $R=21.91$ &
 $\cdot\cdot\cdot$ & $\cdot\cdot\cdot$ & $\cdot\cdot\cdot$\\
\object{MS 2137.3$-$2353} & $14.2$ & $17.8$ & $15.5$ & $V=22.0$ & 
 $0.62$ & $0.41$ & $1.1$\\
\hline\\
\end{tabular}

\noindent
$\star$ From the second brightest cluster member galaxy.
 For further details, see Luppino \& Gioia (\cite{LG92}) and
 Ellingson et al.\ (\cite{Ellingson97}).
\end{table*}
%
%
\begin{table*}
\caption[]{Expected number of giant luminous arcs.
 Case I: E/S0 SED is UV-intermediate NGC3379 SED.
 Case II: E/S0 SED is UV-bright NGC4649 SED.}
\label{Number}
\begin{tabular}{lccc}
\hline\hline\\
Model & Case I & Case II & observed number \\
\hline\\
Isothermal $\beta$ model & $0.031$ & $0.055$ & $5^\star$\\
ENF98-NFW model          & $1.7$   & $3.8$   & $4^\star$\\
\hline\\
\end{tabular}

\noindent
$\star$ See Sec.\ 4.3.
\end{table*}
%
%
\subsection{Numbers of giant luminous arcs}
As noted above, the object \object{MS 1333.3$+$1725} is not a cluster 
and thus was excluded from the cluster sample.
Since the central surface brightness of cluster \object{MS 1621.5$+$2640}
is very close to the background,
X-ray emission from only the central small portion was 
resolved by {\sl HRI}\ (Morris et al.\ \cite{Morris98}).
Because of this reason and poor photon statistics,
the standard $\beta$ model fitting 
gave unusual values as the best-fit result and 
it was impossible to estimate its errors.
The ENF98 $\beta$ model fitting for this cluster did not converge.
The cluster \object{MS 1621.5$+$2640} in which a GLA is found
hence was excluded from the  
arc statistics sample in the current studies,
following which the observed number of GLAs in the sample
of 14 clusters becomes 5.
Although two giant arcs were detected by the Hubble Space Telescope in
clusters \object{MS 1358.4$+$6245} (Franx et al. \cite{Franx97}) and
\object{MS 1512.4$+$3647} (Seitz et al. \cite{Seitz97}),
they are not extremely bright and hence do not enter the
GLA statistics.
X-ray emission from object \object{MS 2053.7$-$0449} was not detected 
in spite of pointing observation with 5 ksec exposure time. 
The $3 \sigma$ upper limit on the X-ray flux of \object{MS 2053.7$-$0449}
was set.
The upper limit on the temperature could be obtained from this
upper limit by following procedure.
First the temperature was assumed to be 6 keV and the upper limit of
X-ray luminosity was calculated. 
Then using the $L_{\rm X}-T$ relation, 
the temperature value was reset
and an upper limit of X-ray luminosity was re-calculated
with this temperature.
Iterating this procedure until the value of the temperature converged,
the upper limit on temperature listed in Table \ref{FLT}
was obtained.
The $\beta$ value of \object{MS2053.7$-$0449} was assumed to be $2/3$.
A core radius of 10 arcsec for \object{MS2053.7$-$0449} was also assumed
because smaller core radius makes a cluster more efficient to make GLAs.
Unfortunately, we could not calculate the central electron number density
of \object{MS2053.7$-$0449} and therefore \object{MS2053.7$-$0449}
was excluded from the GLA statistics with the ENF98-NFW model 
(and with MSS98-NFW when we calculate total cross-section to make
giant arcs).
The expected number of GLAs hence becomes 4/13 in this case.
The expected numbers of GLAs in LF94 sample, $N_{\rm arc}$, were calculated
summing up all the expected numbers of GLAs in each cluster, $N_{i,{\rm arc}}$,
which were described in detail in Paper I.

We list the result in Table \ref{Number}.
It shows that the isothermal $\beta$ model
consistent with current best X-ray data,
cannot reproduce the observed large number of GLAs in the LF94's sample.
On the other hand, the ENF98-NFW model marginally reproduces 
the observed number of GLAs in the LF94 sample.
However, as one can see in Table \ref{ENFNFW},
some clusters have extraordinary high $T_{\rm vir}$
which are in disagreement with the temperatures measured by {\sl ASCA}
or estimated using the $L_{\rm X}-T$ relation of AE98. 
The dominant contribution to increasing
the predicted number of GLAs is these high temperature clusters.

\section{Discussion}

Although the ENF98-NFW model could marginally reproduce 
the observed number of GLAs in the LF94's sample,
some clusters have unrealistic $T_{\rm vir}$.
If ENF98-NFW model describes the truth and 
if it is this high virial temperature that describes the true depth of 
gravitational potential of clusters, 
the non-thermal pressure model (Loeb \& Mao \cite{Loeb94}) or 
the cooling flow model (Allen \cite{Allen98}) would be expected.
However, as shown in Table \ref{Cooling}, cooling times of 
{\it all} the clusters with $T_{\rm vir}> 14$ keV
are longer than the ages of the universe at clusters' redshifts.
This indicates that higher $T_{\rm vir}$ obtained 
from the ENF98 $\beta$ model fitting
were not due to the existence of cooling flows.
The high values of $T_{\rm vir}$ are likely due to too large
values of the scale radii $r_s$.
The large values of scale radii ($r_s$) come from large X-ray core radii
($r_c$) and those who have large X-ray core radii show
significant ellipticity.
As NFW discussed,
their profile came from the virialized system and 
those which have too high $T_{\rm vir}$ hence would be regarded
as in relaxing process.

For the comparison,
we calculated a total cross-section to make giant arcs with another NFW type model
in which temperatures are required to be those 
that are observed by {\sl ASCA} or 
that are estimated using $L_{\rm X}-T$ relation of AE98 
using the result of a theoretical work by Makino et al.\ (\cite{Makino98},
henceforth MSS98).
MSS98 computed the X-ray cluster gas density distribution 
in hydrostatic equilibrium from NFW model 
assuming isothermality of the ICM.
MSS98 showed that the resulting distribution was well approximated by
the standard $\beta$ model.
Their result gives relations of 
$$ r_s = r_c/0.22 \sim 4.55 r_c ,$$
$$ \beta = 0.9 b ,$$
and
$$ b = \frac{8 \pi G \mu m_p \delta_c \rho_{\rm crit}{r_s}^2}
{27 k_{\rm B} T_{\rm X}} ,$$
and we could evaluate $\delta_c$ and $r_s$ (henceforth we call this model 
{\it MSS98-NFW model}).
We list the parameters of MSS98-NFW in Table \ref{ENFNFW} and
we plotted the total cross-section to make giant arcs with
MSS98-NFW model with long-dashed line in Fig.\ \ref{CS}.
As one can see in Fig.\ \ref{CS},
the total cross-section to make giant arcs is considerably small and 
almost the same as the isothermal $\beta$ model.
This means the NFW model which is consistent with the ICM spatial and spectral
data of sample clusters cannot reproduce the observed number of GLAs.

This is also the same to say that what is needed is just to make the sample clusters'
temperatures much higher to reproduce the observed number of GLAs.
However, such high temperatures are no more consistent with ASCA results or
expected values from the $L_{\rm X}-T$ relation of AE98.\\

Systematic errors introduced by uncertainties
in the background galaxy model we employed are summarized as follows.

As is in Paper I,
the luminosity function was taken from Efstathiou et al.\
(\cite{Efstathiou88}). 
This luminosity function is in good agreement with the recent Las Campanas
(Lin et al.\ \cite{Lin96}) and the Stromro-APM (Loveday et al.\
\cite{Loveday92}) redshift surveys.
These are called luminosity functions with lower normalization.
On the other hand,
the ESO Slice Project redshift survey (Zucca et al.\ \cite{Zucca97})
gives higher normalization. 
Its amplitude is then higher, by a factor of $\sim$ 1.6 at $M \sim M^*$.
Luminosity functions with higher normalization thereby increase 
the number of GLAs, with rough estimation, by a factor of two.
As noted above, we assumed no evolution of galaxy number density in
the comoving volume for the luminosity function.
On the other hand,  HF97 showed that observed evolution of the luminosity function, 
which came from Canada France Redshift Survey (Lilly et al.\ \cite{Lilly95}), 
increases the number of GLAs by a factor of several.

We investigated whether changing the type mixing ratio of background galaxies
affected the GLA number by changing (E/S0, Sab, Sbc, Scd, Sdm)=
(0.321, 0.281, 0.291, 0.045, 0.062) and (0.38, 0.16, 0.25, 0.10, 0.11)
and then little difference was found.

As is in Paper I, 
the amplification factor was approximately constant over the whole area of an image 
and the validity of this assumption has already been confirmed in Paper I.

The intrinsic ellipticity of source galaxies
could increase the number of GLAs by a factor of two
as discussed in Paper I.

All these may affect on the expected number of GLAs
by about an order of magnitude at most.
Therefore even if uncertainties in the background galaxy model
are taking into account,
the main conclusion never change;
the models consistent with the ICM spatial and spectral
data of sample clusters, cannot reproduce
the observed number of GLAs in the LF94's sample.\\

As noted in Sec.\ 4, 
recent observations give us new insights on 
galaxy surface brightness and size evolution 
up to $z \sim 4$ (Roche et al. \cite{Roche98} and references therein).
There is no size and luminosity evolution of elliptical galaxies
at higher redshift (Roche et al. \cite{Roche98}).
Spiral galaxies become smaller in size and brighter in surface brightness as 
redshift increases (Roche et al. \cite{Roche98}).
Miralda-Escud\`e (\cite{Miralda93b}) and Paper I showed
that the number of GLAs responds sensitively on the intrinsic size of galaxies
and the number of GLAs decreases drastically 
when the intrinsic size of galaxies becomes smaller than the seeing
FWHM. (See Fig.~3 in Miralda-Escud\`e \cite{Miralda93b} or Fig.~4 in Paper I).
We believe that this effect is stronger than that of surface brightness evolution
because spirals seems not showing strong evolution in luminosity.\\

If the galaxy evolution history is drawn with the merger model
which is currently popular  
(e.g.\ Kauffmann \cite{Kauffmann97}; Bekki \cite{Bekki97}; 
Bekki \& Shioya \cite{Bekki97}; Noguchi \cite{Noguchi97}),
the galaxy evolution model we employed should be largely modified.
Owing to merging-induced star formation,
the merger model predicts the existence of temporarily
very bright galaxies at various redshifts.  
Since the current galaxies are to be formed by aggregation of 
smaller building blocks in the merger model, 
the number density of source galaxies at high redshift 
is larger than the current galaxy number density.
These two effects may thus increase the number of GLAs. 
Although the precise modeling of the merger history
is required to quantify the effect,
we can regard that 
these effects are included as re-normalization in the evolution of the
galaxy luminosity function. 
The result obtained by HF97 thus provided a rough 
idea how this effect changes model prediction.
On the other hand,
as we discussed above,
smaller size of the block-building galaxies at high redshift 
leads drastic decrease of the number of GLAs. 
Becoming larger in size by merging is competed by becoming less in number
of galaxies by merging.
On the arc statistics, the effect of being smaller intrinsically 
seems stronger than both
being more luminous intrinsically and being numerous at high redshift.\\

Asada (\cite{Asada98}) showed that the use of the Deyer-Roeder distance 
to take into account the inhomogeneity of matter distribution in the universe
decreased the cross-section for forming GLAs for all the sets of ($\Omega, \Lambda, H$).
Therefore the appropriate application of angular diameter distance taking account of 
the inhomogeneity of the universe further decrease the number of GLAs.\\

Our calculation assumed that one GLA is generated from one single source galaxy.
However, it may happen that two or more GLAs are generated from a single source galaxy.
This means that the {\it `true'} expected numbers of GLAs exists between
one times the expected numbers we calculated and two times of them
under the spherically symmetric mass distribution models.\\

We close our discussion by this simple question:
are {\it all} observed GLAs in LF94's sample {\it really} giant luminous arcs?
LF94 discussed the possibility of mis-identification by elongated objects
and claimed that
6 GLAs they found were really GLAs
either because their widths were not spatially resolved,
or because they presented a well-defined curvature.
However, the giant luminous arcs in \object{MS 1910.5$+$6736} and \object{MS1008.1-1224}
are located 67 and 47 arcsec 
away from cluster center respectively, which are
unusually large values that would make the cluster extremely massive.
Although, such mis-identification does not affect our result,
it would seriously affect the GLA statistics
which use the number of GLA on the whole sky
extrapolating the GLA detecting rate in EMSS sample.
Spectroscopic conformation of GLAs therefore is needed for all the GLAs in
the sample.

\section{Conclusions}
We studied the statistics of GLAs 
with spherically symmetric lens models based on our original X-ray spatial data
of all the 15 clusters in the LF94's arc survey sample, obtained by {\sl ROSAT HRI}.
We re-investigated whether 
the observed number of GLAs in the LF94's cluster sample
could be reproduced with the cluster mass distribution models 
consistent with the X-ray data of spatial distribution of the ICM
within a frame work of spherical symmetry 
assuming isothermality and the hydrodynamical equilibrium.
We employed two types of cluster mass distribution models.
One is a model comes from the conventional $\beta$ model used for X-ray spatial data fittings
(the 'isothermal $beta$ model' in this paper).
The other is the universal dark matter halo profile model proposed by NFW
(the 'ENF98-NFW model' and the 'MSS98-NFW model' in this paper).
The ENF98-NFW model is a model with the result of cosmological and hydrodynamical
simulations to examine the evolution of X-ray emitting hot gas in clusters, by ENF98.
The MSS98-NFW model is a model with the result of the theoretical work by MSS98.
Models consistent with current best data of spatial distribution of the ICM in the
sample clusters taken by {\sl ROSAT HRI} (isothermal $beta$ model and MSS98-NFW model)
gave the numbers of GLAs less than two orders of magnitude 
and this fewness is significant comparing to the uncertainties in the background galaxy model 
we employed.
On the other hand, ENF98-NFW model almost reproduced the observed number of GLAs.
Some clusters' virial temperatures of this model, however, 
are much higher than the temperatures measured by {\sl ASCA} 
or evaluated from the $L_{\rm X}-T$ relation of AE98.
This indicates that either: non-thermal components of the pressure
play a significant role in supporting the ICM (e.g. Loeb \& Mao \cite{Loeb94}) 
or the mass distribution of sample clusters deviates significantly
from the spherical symmetry.
For example, MS1006.0$+$1202 has straight arcs
which cannot be formed by spherically symmetric mass distribution.
Some sample clusters show significant 
irregularity in their X-ray morphologies.
We believe that 
taking into account the irregularity in mass distribution of clusters
is therefore very important subject and will constitute the next
step together with a better handling of the temperature measurement of
the ICM.
%
%
\begin{acknowledgements}
KM and MH would like to thank Toshifumi Futamase and Takashi Hamana
for insightful comments.
JPK thanks the Astronomical Institute of Tohoku University
and Yamada Science Foundation for fruitful visit to Japan.
This work was supported by the Sasakawa Scientific Research Grant from
the Japan Scientific Society (10-099).
MH is supported by Yamada Science Foundation and by the
Grants-in-Aid by the Ministry of Education, Science and Culture of
Japan (09740169, 60010500). 
\end{acknowledgements}
%
%
%
\section*{Appendix.}
We note down individual clusters with overlays of significance contour plots
superposed on the Digitized Sky Survey
\footnote{The Digitized Sky Surveys were produced
at the Space Telescope Science Institute
under U.S.\ Government grant NAGW-2166.
The images of these surveys are based on photographic data
obtained using the Oschin Schmidt Telescope
on Palomar Mountain and the UK Schmidt Telescope.
The plates were processed into the present compressed digital form
with the permission of these institutions.

The National Geographic Society $-$ Palomar Observatory Sky Atlas (POSS-I)
was made by the California Institute of Technology
with grants from the National Geographic Society. 

The Second Palomar Observatory Sky Survey (POSS-II)
was made by the California Institute of Technology
with funds from the National Science Foundation,
the National Geographic Society,
the Sloan Foundation,
the Samuel Oschin Foundation, and the Eastman Kodak Corporation. 

The Oschin Schmidt Telescope is operated
by the California Institute of Technology and Palomar Observatory. 

The UK Schmidt Telescope was operated
by the Royal Observatory Edinburgh,
with funding from the UK Science and
Engineering Research Council
(later the UK Particle Physics and Astronomy Research Council),
until 1988 June, and
thenceforth by the Anglo-Australian Observatory.
The blue plates of the southern Sky Atlas and its Equatorial Extension
(together known as the SERC-J), as well as the Equatorial Red (ER),
and the Second Epoch [red] Survey (SES) were all taken with the UK Schmidt.}
optical images and radial profiles.
A significance contour plot is made from an X-ray image 
smoothed with a Gaussian filter with a
$\sigma$ of $15\arcsec$ after subtracting the background.
The lowest contour levels is $2 \sigma$.
In each radial profile, the solid line and the dashed line 
respectively represent the result of fitting with standard $\beta$
model and ENF98 $\beta$ model. 
The dotted line represents the point spread function 
at the position of each X-ray center.

\subsection*{\object{MS 0015.9$+$1609} (Cl 0016+16)}
One correlation was found between the brightest point source in the
{\sl HRI} field of view and objects cataloged in ROSID: QSO0015+162.
As Neumann \& B\"ohringer(\cite{Neumann97}, henceforth NB97) mentioned,
there are two observations for this cluster.
Each exposure time was 70 ksec and 2 ksec.
NB97 merged two observational data referring the position of the QSO.
However, since 1)there is only one source available for the pointing check of 
2 dimensional position and 2) 70 ksec exposure time is long enough compared
with $70+2=72$ ksec exposure time, we used only the data of 70 ksec exposure
time to avoid systematic error which comes from merging 2 dimensional 
data referring only one source position.
Our result of standard $\beta$ model fitting to {\sl HRI} data is
slightly different from NB97's result of the $\beta$ model fitting to 
HRI data (but consistent with it within 90 \% errors) and rather consistent 
with their results of that of PSPC data.
The position of QSO0015+162 on {\sl HRI} is 
($00^{\rm h}18^{\rm m}32\fs15$, $+16^{\rm d}29^{\rm m}25\fs5$)(J2000).
The position of X-ray center on {\sl HRI} is
($00^{\rm h}18^{\rm m}33\fs73$, $+16^{\rm d}26^{\rm m}07\fs5$)(J2000)
%
%
\subsection*{\object{MS 0302.7$+$1658} (\object{Cl 0302+1658})}
One observation was performed for this cluster.
No correlation was found
between sources higher than 3 $\sigma$ in the {\sl HRI} field of view 
and objects cataloged in HRASSCAT or ROSID.
The position of the X-ray center on {\sl HRI} is 
($03^{\rm h}05^{\rm m}31\fs94$, $+17^{\rm d}10^{\rm m}05\fs$0)(J2000).
The position of the X-ray peak is consistent with that of 
the brightest cluster member galaxy within the {\sl HRI} pointing accuracy.
%
%
\subsection*{\object{MS 0353.6$-$3642} (\object{S 400})}
One observation was performed for this cluster.
No correlation was found
between sources higher than 3 $\sigma$ in the {\sl HRI} field of view 
and objects cataloged in HRASSCAT or ROSID.
This cluster does not show a significant X-ray peak, i.e. the X-ray center, 
we therefore defined the X-ray center as source count rate weighted mean
position of sources around pointing center.
The source count rates are calculated by the EXSAS command ``DETECT/SOURCES''.
%
%
\subsection*{\object{MS 0451.5$+$0250} (\object{Abel 520})}
One observation was performed for this cluster.
No correlation was found
between the sources higher than 3 $\sigma$ in the {\sl HRI} field of view 
and objects cataloged in HRASSCAT or ROSID.
This cluster does not show a significant X-ray peak, i.e. the X-ray center, 
we therefore define the X-ray center as source count rate weighted mean
position of sources around pointing center.
The source count rates were calculated by the EXSAS command ``DETECT/SOURCES''.
%
%
\subsection*{\object{MS 0735.6$+$7421} (\object{ZwCl 0735.7+7421})}
One observation was performed for this cluster.
No correlation was found
between sources higher than 3 $\sigma$ in the {\sl HRI} field of view 
and objects cataloged in HRASSCAT or ROSID.
The position of the X-ray center is consistent with that of 
the brightest cluster member galaxy within the {\sl HRI} pointing accuracy.
%
%
\subsection*{\object{MS 1006.0$+$1202} (\object{ZwCl 1006.1+1201})}
Three observations were performed for this cluster.
Each exposure time was 3 ksec, 20 ksec, and 65 ksec.
No correlation was found
between sources higher than 3 $\sigma$ in the {\sl HRI} field of view 
and objects cataloged in HRASSCAT or ROSID.
We used only the data of 65 ksec exposure time 
to avoid systematic error in merging the data without any references
because 65 ksec exposure seemed long enough.
This cluster does not show a significant X-ray peak, i.e. the X-ray center, 
we therefore defined the X-ray center as source count rate weighted mean
position of sources around pointing center.
The source count rates were calculated by the EXSAS command ``DETECT/SOURCES''.
%
%
\subsection*{\object{MS 1008.1$-$1224}}
One observation was performed for this cluster.
No correlation was found
between sources higher than 3 $\sigma$ in the {\sl HRI} field of view 
and objects cataloged in HRASSCAT or ROSID.
X-ray center (J2000.0) is
($10^{\rm h}10^{\rm m}32\fs42$, $-12^{\rm d}39^{\rm m}47\fs0$).
%
%
\subsection*{\object{MS 1224.7$+$2007}}
Four observations were performed for this cluster.
Each exposure time was 14 ksec, 24 ksec, 5 ksec, and 13 ksec.
No correlation was found
between sources higher than 3 $\sigma$ in the {\sl HRI} field of view 
and objects cataloged in both HRASSCAT and ROSID.
One, and the only one, cataloged bright point source was available
for comparison of pointing accuracy in each image.
We therefore used only the data of 24 ksec exposure time which was the
longest one among them, 
to avoid systematic error 
which came from merging data referring only one source position.
X-ray center (J2000.0) is 
($12^{\rm h}27^{\rm m}13\fs28$, $+19^{\rm d}50^{\rm m}57\fs$0).
%
%
\subsection*{\object{MS 1333.3$+$1725}}
One observation was performed for this cluster.
One correlation was found
between sources higher than 3 $\sigma$ in the {\sl HRI} field of view 
and objects cataloged in HRASSCAT and ROSID: QSO1333+177.

As the contour plot and the radial profile show,
this object is not a cluster 
but an X-ray point source.
Therefore this object was excluded from the sample.
%
%
\subsection*{\object{MS 1358.4$+$6245} (\object{ZwCl 1358.1+6245})}
Two observations ware performed for this cluster.
Each exposure time was 14 ksec and 16 ksec.
One correlation was found
between sources higher than 3 $\sigma$ in the {\sl HRI} field of view 
and objects cataloged in HRASSCAT: J135903.4+621239.
However, J135903.4+621239 locates near the edge of {\sl HRI} field of view
and was not usable for the reference of the pointing accuracy check.
There was also a bright source not cataloged
and thus only one point source near the pointing center 
available for the reference of the pointing accuracy check.
Since the exposure times of two observations were comparable to each other 
and each exposure time is not long enough,
we chose the way to merge two files believing that pointing of each
observation is completely identical.
The difference of position of the point source near the pointing center
is $3\arcsec$ in RA and $2\arcsec$ in DEC and therefore we can say
that these two positions are identical within the {\sl HRI} pointing accuracy.
Position of the X-ray center on {\sl HRI} is
($13^{\rm h}59^{\rm m}50\fs34$, $+62^{\rm d}31^{\rm m}04\fs5$)(J2000).
The position of the X-ray center is consistent with that of 
the brightest cluster member galaxy within the {\sl HRI} pointing accuracy.
%
%
\subsection*{\object{MS 1455.0$-$2232} (\object{1E 1455+2232})}
Three observations were performed for this cluster.
Each exposure time was 4 ksec, 4 ksec, and 7 ksec.
No correlation was found
between sources higher than 3 $\sigma$ in the {\sl HRI} field of view 
and objects cataloged in HRASSCAT or ROSID.
Since the exposure times of three observations were comparable to each other 
and each exposure time is not long enough,
we chose the way to merge three files believing that the pointing of each
observation was completely identical.
The position of the X-ray center on {\sl HRI} is 
($14^{\rm h}57^{\rm m}14\fs94$, $+22^{\rm d}20^{\rm m}35\fs5$)(J2000).
The position of the X-ray center is consistent with that of 
the brightest cluster member galaxy within the {\sl HRI} pointing accuracy.
%
%
\subsection*{\object{MS 1512.4$+$3647}}
One observation was performed for this cluster.
Two correlations were found
between sources higher than 3 $\sigma$ in the {\sl HRI} field of view 
and objects cataloged in ROSID: QSO1512+370 and HD135657.
The position of the X-ray center on {\sl HRI} is 
($15^{\rm h}14^{\rm m}22\fs65$, $+36^{\rm d}36^{\rm m}20\fs0$)(J2000).
The position of the X-ray center is consistent with that of 
the brightest cluster member galaxy within the {\sl HRI} pointing accuracy.
%
%
\subsection*{\object{MS 1621.5$+$2640}}
One observation was performed for this cluster.
No correlation was found
between sources higher than 3 $\sigma$ in the {\sl HRI} field of view 
and objects cataloged in HRASSCAT or ROSID.
The {\sl HRI}\ image of \object{MS 1621.5$+$2640}
shows double peaks and therefore we defined the X-ray center
as the position of the second brightest cluster member galaxy.
(First brightest cluster member galaxy locates about $3\arcmin$ away
from X-ray center (Ellingson et al. \cite{Ellingson97}). )
This position is near the center of double X-ray peaks.
This cluster shows very poor signal-to-noise ratio.
Forcing standard $\beta$ model fitting gave unusual values for its best-fit
(Morris et al. \cite{Morris98}).
We also fit the {\sl PSPC}\ data, which has the lower background than
{\sl HRI}, 
and only to give us unacceptable $\chi^2$ value 
same as the case of {\sl HRI}.
{\sl PSPC}\ data showed that the position
where the tail of radial profile from {\sl HRI}\
meets the background was still in the cluster source region on the {\sl PSPC}\  image. 
Therefore $\beta$ model fitting to {\sl HRI} data is not applicable.
This cluster is excluded from the arc statistics.
%
%
\subsection*{\object{MS 1910.5$+$6736}}
One observation was performed for this cluster.
No correlation was found
between sources higher than 3 $\sigma$ in the {\sl HRI} field of view 
and objects cataloged in HRASSCAT or ROSID.
The position of the X-ray center on {\sl HRI} is 
($19^{\rm h}10^{\rm m}27\fs94$, $+67^{\rm d}41^{\rm m}28\fs$0)(J2000).
%
%
\subsection*{\object{MS 2053.7$-$0449}}
One observation was performed for this cluster.
No correlation was found
between sources higher than 3 $\sigma$ in the {\sl HRI} field of view 
and objects cataloged in HRASSCAT or ROSID.

In spite of 5 ksec pointing, we could have no source photons of 
\object{MS 2053.7$-$0449}.
This indicates the low signal-to-noise ratio of this cluster.
We gave the upper limit of the central surface brightness assuming 
$\beta_{\rm fit} = 2/3$ for analytical calculation.
This cluster was excluded from our GLA statistics with the ENF98-NFW model
and the MSS98-NFW model.
For the GLA statistics with Isothermal $\beta$ model, core radius was assumed
to be $10\arcsec$.
%
%
\subsection*{\object{MS 2137.3$-$2353}}
Two observations were performed for this cluster.
Each exposure time was 2 ksec and 14 ksec.
No correlation was found
between sources higher than 3 $\sigma$ in the {\sl HRI} field of view 
and objects cataloged in HRASSCAT or ROSID.
Since 14 ksec exposure time is long enough 
comparing with $14+2=16$ ksec exposure time, 
we used only the data of 14 ksec exposure time 
to avoid systematic error which came from merging data 
with no references.
The position of the X-ray center on {\sl HRI} is 
($21^{\rm h}40^{\rm m}15\fs17$, $-23^{\rm d}39^{\rm m}41\fs0$)(J2000).
The position of the X-ray center is consistent with that of 
the brightest cluster member galaxy within the {\sl HRI} pointing accuracy.
%
%
%


\begin{thebibliography}{}
\bibitem[1998]{Allen98}
Allen, S.W., 1998, MNRAS 296, 392
\bibitem[1996]{Allen96}
Allen, S.W., Fabian, A.C., Kneib, J.-P., 1996, MNRAS 279, 615
\bibitem[1986]{Arimoto86}
Arimoto, N., Yoshii, Y., 1986, A\&A 164, 260
\bibitem[1987]{Arimoto87}
Arimoto, N., Yoshii, Y., 1987, A\&A 173, 23
\bibitem[1998]{AE98}
Arnaud, M., Evrard, A.E. 1998, MNRAS submitted.; astro-ph/9806353 (AE98)      
\bibitem[1998]{Asada98}
Asada, H. 1998, ApJ in press;astro-ph/9803004
\bibitem[1994]{Bahcall94}
Bahcall, N.A., Lubin., R.M., 1994, ApJ, 426, 513 
\bibitem[1994]{Bartelmann94}
Bartelmann, M., Weiss, A., 1994, A\&A 287, 1
\bibitem[1995]{Bartelmann95}
Bartelmann, M., Steinmetz, M., Weiss, A. 1995, A\&A 297, 1 (BSW95)
\bibitem[1996]{B96}
Bartelmann, M., 1996, A\&A 313, 697
\bibitem[1996]{Bartelmann96}
Bartelmann, M., Steinmetz M., 1996, MNRAS 283, 431
\bibitem[1998]{Bartelmann98}
Bartelmann, M., Huss, A., Colberg, J.M., Jenkins, A., Pearce, F.R., 1998, 
A \& A, 330, 1
\bibitem[1997]{Bekki97}
Bekki, K., 1997, 
in {\it Galaxy Interactions at Low and High Redshift},
IAU Symposium 186
\bibitem[1998]{Bekki98}
Bekki, K., Shioya, Y., 1998, ApJL 478, L17
\bibitem[1998]{Bezecourt98}
B\`ezecourt, J., Pell\'o, R., Soucail, G., A\&A 330, 399
 \bibitem[1998]{Boehringer98}
B\"ohringer, H., Tanaka, Y., Mushotzky, R.F., Ikebe, Y., Hattori, M., 1998, A\&A 334, 789
\bibitem[1976]{Cavaliere76}
Cavaliere, A., Fusco-Femiano, R., 1976, A\&A, 49, 137
\bibitem[1993]{David93}
David, L.P., Slyz, A., Jones, C., Forman, W., Virtlek, S., Arnaud, K.,
1993, ApJ, 412, 479	            
\bibitem[1990]{DL90}
Dickey, J.M., Lockman, F.J. 1990, ARA\&A, 28, 215	      
\bibitem[1988]{Efstathiou88}
Efstathiou, G., Ellis, R.S., Peterson, B.A., 1988, MNRAS 232, 431
\bibitem[1997]{ENF98}
Eke, V.R., Navarro, J.F., Frenk, C.S., 1998, ApJ 503, 569
\bibitem[1997]{Ellingson97}
Ellingson, E., Yee, H.K.E., Abraham, R.G., Morris, S.L., Carlberg, R.G.,
Smecker-Hane, T.A., 1997, ApJS 113, 1      
\bibitem[1998]{Ellis98}
Ellis, R., 1998,
in {\it Cosmological Parameters \& Evolution of the Universe},
ed Sato, K., IAU Symposium 183
\bibitem[1994]{Fabian94}
Fabian, A.C., 1994, ARA\&A 32, 277	      
\bibitem[1994]{Fort94}
Fort, B., Mellier, Y., 1994, A\&AR 5, 239
\bibitem[1997]{Franx97}
Franx, M., Illingworth, G.D., Kelson, D.D., Van Dokk Um, P.G., Tran,
	      K.-V., 1997 ApJL, 486, L75
\bibitem[1997]{Hamana97}
Hamana, T., Futamase, T., 1997, MNRAS 286, L7 (HF97)
\bibitem[1991]{Hammer91}
Hammer, F., 1991, ApJ 383, 66
\bibitem[1999]{Hattori99}
Hattori, M., Kenib, J.-P., Makino, N., 1999, Progress of Theoretical Physics, Supplement, 133, 1
\bibitem[1998]{Hattori98}
Hattori, M., Matuszawa, H., Morikawa, K., Kneib, J.-P., Yamashita, K.,
Watanabe, K., B\"ohringer, H., Tsuru, T.G., 1998, ApJ in press;
astro-ph/9803092
\bibitem[1997]{Hattori97}
Hattori, M., Watanabe, K., Yamashita, K. 1997, A\&A 319, 764 (Paper I)
\bibitem[1997]{Henry97}
Henry, J.P., 1997, ApJ 489, L1
\bibitem[1997]{HH86}
Henry, J.P., Henriksen, H.J. 1986, ApJ 301, 689
\bibitem[1997]{Kauffmann97}      
Kauffmann, G., 1997, 
in {\it Galaxy Interactions at Low and High Redshift},
IAU Symposium 186
\bibitem[1991]{Kim91}
Kim, K.-T., Tribble, P.C., Kronberg, P.P., 1991, ApJ 379, 80
\bibitem[1962]{King62}
King, I., 1962, ApJL 174, L123	      
\bibitem[1996]{Kneib96}
Kneib, J.-P., Ellis, R.S., Smail, I., Couch, W.J., Sharples, R.M.,
ApJ 471, 643
\bibitem[1993]{Kneib93}
Kneib, J.-P., Mellier, Y., Fort, B., Mathez, G., 1993, A\&A 273, 367
\bibitem[1995]{Kneib95}
Kneib, J.-P., Mellier, Y., Pell\'o, R., Miralda-Escud\'e, J.,
Le Borgne, J.-F., B\"ohringer, H., Picat, J.-P., 1995 A\&A 303, 27
\bibitem[1994]{LF94}
Le F\`evre, O., Hammer, F., Angonin, M.C., Gioia, I.M., Luppino, G.A., 1994, ApJL 422, L5 (LF94)
\bibitem[1995]{Lilly95}
Lilly, S.J., Tresse, L., Hammer, F., Crampton, D., Le F\`evre, O., 1995, 
ApJ 455, 108
\bibitem[1996]{Lin96}
Lin, H., Kirshner, R.P., Shectman, S.A., Landy, S.D., Oemler, A.,
Tucker, D.L.Schechter P.L., 1996, ApJ 464, 60
\bibitem[1994]{Loeb94}
Loeb, A., Mao, S., 1994, ApJL 435, L17
\bibitem[1992]{Loveday92}
Loveday, J., Peterson, B.A., Efstahiou, G., Maddox, S.J. 1992, ApJ 390, 338
\bibitem[1993]{Lubin93}
Lubin, L.M., Bahcall,, N.A., 1993, ApJL 415, L17
\bibitem[1992]{LG92}
Luppino, G.A., Gioia, I.M., 1992, A\&A 265, L9
\bibitem[1986]{Lynds86}
Lynds, R., Petrosian V. 1986, BAAS 18,1014
\bibitem[1996]{Madau96}
Madau, P., Ferguson, H.C., Dickinson, M.E., Giavalisco, M., Steidel,
C.C., Fruchter, A. 1996, MNRAS 283, 1388
\bibitem[1998]{Makino98}
Makino, N., Sasaki, S.,  Suto Y., 1998, ApJ 497, 555
\bibitem[1998]{Markevitch98}
Markevitch, M., 1998, ApJ submitted; astro-ph/9802059
\bibitem[1993a]{Miralda93a}
Miralda-Escud\'e, J., 1993a, ApJ 403, 497
\bibitem[1993b]{Miralda93b}
Miralda-Escud\'e, J., 1993b, ApJ 403, 509
\bibitem[1995]{Miralda95}
Miralda-Escud\'e, J., Babul, A., 1995, ApJ 449, 18
\bibitem[1998]{Morris98}
Morris, S.L., Hutchings, J.B., Carlberg, R.G., Yee, H.K.C., Ellingson,E., 
Balogh, M.L., Abraham, R.G., Smecker-Hane T.A., 1998,
ApJ in press;astro-ph/9805216
\bibitem[1997]{Mushotzky97}
Mushotzky, R.F., Scharf, C.A., 1997, ApJL 482, L13
\bibitem[1995]{Narayan95}
Narayan, R., Bartelmann, M., in {\it Formation of Structure in
theUniverse}, Proceedings of the 1995 Jerusalem Winter School,
ed A. Dekel and J.P. Ostriker (Cambridge University Press)
\bibitem[1996]{Navarro96}
Navarro, J.F., Frenk, C.S., White, S.D.M., 1996, ApJ 462, 563 (NFW)
\bibitem[1997]{Navarro97}
Navarro, J.F., Frenk, C.S., White, S.D.M., 1997, ApJ 490, 493 (NFW)
\bibitem[1997]{Neumann97}
Neumann, D.M., B\"ohringer, H., 1997, MNRAS 289, 123 (NB97)
\bibitem[1997]{Noguchi97}
Noguchi, M., 1997, 
in {\it Galaxy Interactions at Low and High Redshift},
IAU Symposium 186      
\bibitem[1976]{Pence76}
Pence, W., 1976, ApJ 203, 39
\bibitem[1996]{Pierre96}
Pierre, M., Le Borgne, J.F., Soucail, G., Kneib, J.-P., 1996, A\&A 311,	413
\bibitem[1998]{Roche98}
Roche, N., Ratnatunga, K., Griffiths, R.E., Im, M., Naim, A., 
1998, MNRAS 293, 157
\bibitem[1997]{Seitz97}
Seitz, S., Saglia, R.P., Bender, R., Hopp, U., Belloni, P., Ziegler, B., 1997,
MNRAS submitted; astro-ph/9706023
\bibitem[1987]{Soucail87}
Soucail, G., Fort, B., Mellier, Y., Picat, J.P., 1987, A\&A 172, L14
\bibitem[1988]{Soucail88}
Soucail, G., Mellier, Y., Fort, B., Mathez, G., Cailloux, M., 1988, A\&A 191, L19
\bibitem[1997]{Totani97}
Totani, T., Yoshii, T., Sato, K. 1997, ApJ 483, L75
\bibitem[1984]{Trumper84}
Tr\"umper, J., 1984, Physica Scripta 7, 209
\bibitem[1999]{Umetsu99}
Umetsu, K., Tada, M., Futamase, T., 1999, Progress of Theoretical Physics, Supplement, 133
\bibitem[1993]{Wu93}
Wu, X.-P., Hammer, F., 1993, MNRAS 262, 187
\bibitem[1988]{Yoshii88}
Yoshii, Y., Takahara, F., 1988, ApJ 326, 1
\bibitem[1991]{Yoshii91}
Yoshii, Y., Perterson B.A., 1991, ApJ 372, 8
\bibitem[1997]{Zucca97}
Zucca, E., Zamorani, G., Vettolani, G., Cappi, A., Merighi, R., Mignoli, M., 
Stirpe, G.M., MacGillivray, H.\ et al., 1997 A\&A 326, 477
\end{thebibliography}
\end{document}